\documentclass[11pt,a4paper]{article}

\usepackage[a4paper,text={150mm,240mm},centering,headsep=10mm,footskip=15mm]{geometry}
\usepackage[T1]{fontenc}
\usepackage{epsfig, float,color}
\usepackage{amsmath,amssymb, amscd}
\usepackage{amsfonts,amsbsy}
\usepackage{hyperref}
\usepackage{url}
\usepackage{bbm}
\usepackage{mathrsfs} 
\usepackage{cite}
\usepackage{dsfont}

\newcommand{\R}{\mathbb{R}}

\newcommand{\N}{\mathbb{N}}

\newcommand{\M}{\mathbb{M}}
\newcommand{\id}{\mathbbm{1}}

\newcommand{\diag}{{\rm diag} }
\newcommand{\vx}{{\mathbf{x}} }

\newcommand{\Banach}{\mathscr{B}}
\newcommand{\spacelike}{\mathscr{S}}

\newcommand{\ret}{{\rm ret}}
\newcommand{\adv}{{\rm adv}}
\newcommand{\sym}{{\rm sym}}
\newcommand{\free}{{\rm free}}

\newcommand{\be}{\begin{equation}}
\newcommand{\ee}{\end{equation}}

\newcommand{\qed}{\nobreak \ifvmode \relax \else
      \ifdim\lastskip<1.5em \hskip-\lastskip
      \hskip1.5em plus0em minus0.5em \fi \nobreak
      \vrule height0.75em width0.5em depth0.25em\fi}

\title{Direct interaction along light cones\\ at the quantum level}

\author{
Matthias Lienert\thanks{lienertmat@gmail.com, Department of Mathematics, Rutgers University, 110 Freylinghuysen Road, Piscataway, NJ 08854-8019, USA}
}

\date{October 8, 2018}

\begin{document}

\maketitle

  \begin{abstract}
  	\noindent Here, we point out that interactions with time delay can be described at the quantum level using a multi-time wave function $\psi(x_1,...,x_N)$, i.e., a wave function depending on one spacetime variable $x_i = (t_i,\vx_i)$ per particle. In particular, such a wave function makes it possible to implement direct interaction along light cones (not mediated by fields), as in the Wheeler-Feynman formulation of electrodynamics. Our results are as follows. (1) We derive a covariant two-particle integral equation and discuss it in detail. (2) It is shown how this integral equation (or equivalently, a system of two integro-differential equations) can be understood as defining the time evolution of $\psi$ in a consistent way. (3) We demonstrate that the equation has strong analogies with Wheeler-Feynman electrodynamics and therefore suggests a possible new quantization of that theory. (4) We propose two natural ways how the two-particle equation can be extended to $N$ particles. It is shown that exactly one of them leads to the usual  Schr\"odinger equation with Coulomb-type pair potentials if time delay effects are neglected.
  	
  	\vspace{0.3cm}
  	
  \end{abstract}

\section{Introduction} \label{sec:intro}

In classical physics, electromagnetic interactions can be expressed in two empirically equivalent but conceptually and mathematically very different ways: 
\begin{enumerate}
	\item \emph{Interaction mediated by electromagnetic fields:} the well-known Maxwell-Lorentz electrodynamics, and
	\item \emph{Direct interaction of world lines along light cones:} "action-at-a-distance-electrodyn\-amics" or "Wheeler-Feynman electrodynamics" (WF), see \cite{wf1,wf2} and references therein.
\end{enumerate}
In quantum physics, the first approach corresponds to the second-quantized electromagnetic field of quantum electrodynamics. Relatively little, however, has been said about a quantum analog of the second possibility, direct electromagnetic interactions.\footnote{See Feynman's Nobel Lecture \cite{feynman_nobel_lecture} as well as references \cite{davies_1970,davies_1971,davies_1972,hoyle_narlikar,barut_multiparticle} and \cite[chap. 8]{dirk_phd}.}

The goal of this paper is to introduce a new approach in this second direction. For simplicity, we study the case of $N=2$ interacting particles (electrons) first. A crucial feature of direct electromagnetic interactions is that the motion of a particle at a spacetime point $x$ depends on the motion of other particles at points $x'$ which lie on the light cone of $x$. That means, interaction happens \textit{with a time delay}. It is a major question how to express such a time delay at the quantum level.

In this regard, wave functions $\psi(x_1,...,x_N)$, where $x_i = (t_i,\vx_i)$ are spacetime points,\footnote{We set $\hbar = c = 1$ and use the metric signature + -- -- --.} turn out very useful. These \textit{multi-time wave functions} were first suggested as a basic concept in relativistic quantum theory by Dirac in 1932 \cite{dirac_32}, played an important role in the famous works of Tomonaga \cite{tomonaga} and Schwinger \cite{schwinger} and have recently been developed considerably so as to constitute a relativistic version of the  Schr{\"o}dinger picture \cite{nogo_potentials,qftmultitime,multitime_pair_creation,1d_model,nt_model,phd_lienert,2bd_current_cons,deckert_nickel_2016,lpt_2017a,lpt_2017b}. Sec. \ref{sec:multitime} contains a brief review. The crucial point for the present work is the following. Multi-time wave functions include one time variable per particle and therefore allow to use spacetime configurations involving different times, such as the light-like two-particle configuration $((t,\vx), (t-|\vx-\vx'|,\vx') ) \in \M \times \M$ where $\M$ denotes Minkowski spacetime. Exactly this freedom, which the wave function $\varphi(t,\vx_1,...,\vx_N)$ of the usual  Schr{\"o}dinger picture lacks, makes it possible to express interactions with time delay at the quantum level.

This leads us to consider new possibilities for evolution equations for $\psi$ such as integral equations (similar to the Bethe-Salpeter equation \cite{bs_equation,nakanishi}) as well as integro-differential equations.
The crucial feature of these evolution equations is that they contain integrals over light-like configurations or even more general spacetime configurations. Because of this unusual mathematical structure, several important questions arise.
\begin{enumerate}
	\item Can we guess evolution equations expressing direct interaction with time delay?
	\item What is the general mathematical structure of these equations? When are they mathematically consistent?
	\item What are suitable initial data? In other words: Which part of the solution can be "freely chosen" to determine the rest?
	\item Can we see in more detail that the new evolution equations correspond to a quantum version of WF electrodynamics?
\end{enumerate}

We shall address these questions as follows. 
First, we give an overview of the theory of multi-time wave functions (Sec. \ref{sec:multitime}). Second, we present a heuristic derivation of a covariant two-particle integral equation for a multi-time wave function (Sec. \ref{sec:derivation}). The main features of the equation are discussed and some generalizations are made (Sec. \ref{sec:features}). We also include a brief comparison with the Bethe-Salpeter equation (Sec. \ref{sec:bs}).
Third, we analyze how the integral equation determines the time evolution of the wave function (Sec. \ref{sec:evollaw}). By reformulating the integral equation as a system of integro-differential equations (Sec. \ref{sec:integrodiff}), we make contact with the usual evolution equations for multi-time wave functions. This allows to discuss the consistency of the equations in a similar way as usual for multi-time wave functions (Sec. \ref{sec:integrodiff} and Appendix \ref{sec:ccret}). Fourth, we show that our equations reduce to the usual Schr{\"o}dinger equation with a Coulomb-type potential at equal times in a certain frame when time delay effects are neglected (Sec. \ref{sec:nonrellimit}). Fifth, we discuss the analogies of our equations with WF electromagnetism (Sec. \ref{sec:wf}). Sixth, we present two natural ways to extend the approach to $N \geq 3$ particles: (a) via pair interaction terms (Sec. \ref{sec:npartsums}) and (b) via $N$-particle interaction terms (Sec. \ref{sec:npartint}). For both approaches, we identify concrete evolution equations that reduce to the usual  Schr{\"o}dinger equation with a Coulomb-type potential in the non-relativistic limit.
We conclude with a discussion of the results and give an outlook of possible future directions in Sec. \ref{sec:discussion}.

\section{Multi-time wave functions} \label{sec:multitime}

Multi-time wave functions (see \cite{lpt_2017a} for a more detailed review) arise from the desire to obtain a covariant particle-position representation of a many-particle quantum state.\footnote{One should note that Dirac's multi-time formalism is distinct from the two-time formalism of Aharonov et al. (see \cite{twotime_formalism}) which considers the possibility of prescribing, in addition to the initial state, a final state, either as a result of post-selection or due to an intrinsic boundary condition of the universe.} A multi-time wave function is defined as a scalar or spinor-valued wave function $\psi(x_1,...,x_N)$ depending on $N$ spacetime points $x_i \in \M$  for $N$ particles.\footnote{Particle creation and annihilation can be described using "multi-time Fock functions" $\psi = (\psi^{(0)}, \psi^{(1)}, \psi^{(2)}, \cdots )$ where each $\psi^{(N)}$ is an $N$-particle multi-time wave function (see \cite{qftmultitime,multitime_pair_creation}).} $\psi$ is straightforwardly related to the usual single-time wave function $\varphi$ of  Schr{\"o}dinger's equation by evaluation at equal times (relative to a Lorentz frame):
\be
	\varphi(t,\vx_1,...,\vx_N) = \psi  \!\left( (t,\vx_1), ... , (t,\vx_N) \right).
	\label{eq:singlemulti}
\ee
In contrast with $\varphi$, whose transformation behavior under a Poincar\'{e} transformation $x \mapsto \Lambda x + a$ cannot easily be defined as its argument $(t,\vx_1,...,\vx_N)$ is not a collection of four-vectors, the transformation behavior of a multi-time wave function $\psi$ can readily be stated:
\be
	\psi(x_1,...,x_N) ~\longmapsto~ S_N[\Lambda]\,  \psi \big( \Lambda^{-1} (x_1 - a),...,\Lambda^{-1}(x_N - a) \big),
\ee
where $S_N[\Lambda]$ is a matrix depending on $\Lambda$, for example $S_N[\Lambda] = S[\Lambda]^{\otimes N}$ for $N$ Dirac particles, where the matrices $S[\Lambda]$ form the spinor representation of the Lorentz group.

Eq.\@ \eqref{eq:singlemulti} shows how to obtain a single-time wave function from a multi-time wave function. Conversely, we can start with a single-time wave function $\varphi$, note its implicit reference to a particular frame, and demand that we can Lorentz transform its argument $(t,\vx_1,...,\vx_N)$. To do this, we note the latter refers to the spacetime configuration $((t,\vx_1),...,(t,\vx_N))$ which can be Lorentz transformed to yield $(\Lambda(t,\vx_1),...,\Lambda(t,\vx_N)) = ((t_1',\vx_1'),...,(t_N',\vx_N'))$ where in general $t_i'\neq t_j'$. In this way we see the \textit{necessity} to consider a multi-time wave function if a particle-position representation is desired.

Admitting the most general notion of a configuration of $N$ particles, we are then led to regard the set $\spacelike$ of space-like configurations (or its closure $\overline{\spacelike}$) as the natural domain of $\psi$. In symbols:
\be
	\spacelike = \{ (x_1,...,x_N) \in \M^N : (x_i-x_j)^\mu (x_i-x_j)_\mu < 0~\forall i,j\}.
	\label{eq:spacelikeconfigs}
\ee
In addition to multi-time wave functions on $\spacelike$, multi-time wave functions on $\M^N$ can also be considered. However, this would seem much more radical as $q\in \M^N\backslash \overline{\spacelike}$ cannot be regarded as a configuration of particles.

As $\psi$ contains many time variables, the question of suitable evolution equations arises. Most commonly, so-called \textit{multi-time  Schr{\"o}dinger equations} are considered, i.e., a system of first-order partial differential equations of Hamiltonian form, one for each time variable:
\begin{eqnarray}
	i \partial_{t_1} \psi &=& H_1(t_1,...,t_N) \psi,\nonumber\\
	&\vdots& \label{eq:multitimeeqs} \\
	i \partial_{t_N} \psi &=& H_N(t_1,...,t_N) \psi. \nonumber
\end{eqnarray}
The $H_k$ are called \textit{partial Hamiltonians}. One can think of them as (possibly time-dependent) differential operators on a suitable functions space such as $L^3(\R^{3N})$. The time-dependence is natural if one imagines that covariant potentials could be included in $H_k(t_1,...,t_N)$.

It is understood that the multi-time equations \eqref{eq:multitimeeqs} can be rewritten in a manifestly covariant way, such as in the example of free (non-interacting) multi-time Dirac equations
\be
	(-i \gamma^\mu_k \partial_{k,\mu}+ m_k) \psi = 0,~~~k=1,...,N.
\ee 
Here, $\partial_{k,\mu} = \frac{\partial}{\partial x_k^\mu}$ and $\gamma^\mu_k$ stands for the $\mu$-th Dirac gamma matrix acting on the spin index of the $k$-th particle.

We can obtain a single-time Schr{\"o}dinger equation for $\varphi$ by taking the time derivative of  \eqref{eq:singlemulti} and then using  \eqref{eq:multitimeeqs} and the chain rule. This leads to:
\be
	i \partial_t \varphi = \sum_{k=1}^N H_k \varphi.
	\label{eq:summedschroed}
\ee
Note that values of $\psi$ with different times do not occur in the equation; i.e., we arrive at an autonomous evolution equation for $\varphi$. We emphasize that this is a feature of differential evolution equations (as compared to, for example, integro-differential equations).

The role of the multi-time  Schr{\"o}dinger eqs. \eqref{eq:multitimeeqs} is to determine $\psi$ on $\spacelike$ (or possibly even $\M^N$) from initial data $\psi_0$ on a Cauchy surface $\Sigma$, e.g., at $t_1 = \cdots = t_N = 0$:
\be
	\psi  \!\left( (0,\vx_1), ... , (0,\vx_N) \right) = \psi_0(\vx_1,...,\vx_N).
\ee

The system \eqref{eq:multitimeeqs} contains many simultaneous equations for each component of $\psi$ which must be consistent with each other. It is required that the evolutions in the different time variables commute. This is the case if and only if \cite{bloch,nogo_potentials}:
\be
	[H_j,H_k] - i \frac{\partial H_k}{\partial t_j} + i\frac{\partial H_j}{\partial t_k} = 0~~~\forall j, k.
	\label{eq:cc}
\ee
These \textit{consistency conditions} turn out to be very restrictive. For example, they have been shown to rule out potentials in the following sense  \cite{nogo_potentials,deckert_nickel_2016}. If $H_k = H_k^\free + V_k(x_1,...,x_N)$ where $H_k^\free$ is the free Dirac Hamiltonian acting on the coordinates $x_k$ and spin index of the $k$-th particle, then the only Poincar\'{e}-invariant $V_k$ which satisfy \eqref{eq:cc} are $V_k = {\rm const}$.

Therefore, the need for different mechanisms of interaction for multi-time wave functions arises. In this regard, the following ideas have been analyzed. In \cite{1d_model,nt_model}, relativistic zero-range interactions were introduced and rigorously shown to be consistent for Dirac particles in 1+1 spacetime dimensions. In \cite{qftmultitime,multitime_pair_creation}, the multi-time approach was extended to quantum field theory and particle creation and annihilation were incorporated into the framework. This leads to a mechanism of interaction that satisfies the consistency condition \eqref{eq:cc}; however, one encounters the usual ultraviolet divergences of quantum field theory. This necessitates to introduce a cutoff or to employ renormalization methods. Both these possibilities have been extensively studied in various contexts before, and many fruitful ideas have resulted from them. At the same time, they lead away from a theory based only on a simple, clear, Poincar\'{e}-invariant, fundamental equation and may, therefore, not appear entirely satisfactory from a conceptual standpoint. In view of these limitations and difficulties of the existing approaches, it seems worthwhile to search for novel mechanisms of interaction, especially mechanisms that are made possible or are put forward specifically by the multi-time picture. This constitutes another basic motivation of the present work.

Finally, it is important to note that like the non-relativistic wave function, a multi-time wave functions often carries a physical meaning as a probability amplitude for particle detection. This had long been conjectured by Bloch in \cite{bloch}; a general proof was only given recently in \cite{generalized_born}. There it was shown that assuming the Born rule as well as suitable measurement postulates only in a particular frame leads, for a dynamics with local interactions, to a \textit{Born rule on general Cauchy surfaces} $\Sigma$. That is, a suitable quadratic expression in the multi-time wave function, $|\psi|_\Sigma^2(x_1,...,x_N)$, yields the probability density to detect $N$ particles, at $x_1,...,x_N \in \Sigma$. For example, for $N$ Dirac particles one has:
\be
	|\psi|_\Sigma^2(x_1,...,x_N) = \overline{\psi}(x_1,...,x_N) \gamma \cdot n(x_1) \otimes \cdots \otimes \gamma \cdot n(x_N) \psi(x_1,...,x_N),
	\label{eq:bornrule}
\ee
where $n(x)$ is the normal vector field at $\Sigma$.

\section{A two-particle integral equation with direct interaction along light cones} \label{sec:twopart}

\subsection{Derivation of the equation} \label{sec:derivation}

We now present a heuristic derivation of a promising two-particle equation for a multi-time wave function that expresses direct interaction along light-like configurations. To this end, we start from a reformulation of the single-particle  Schr{\"o}dinger equation as an integral equation. Extending this equation to the two-particle case with Poincar\'{e} invariance in mind naturally leads to an integral equation for a multi-time wave function.

Consider the single-particle  Schr{\"o}dinger equation for the wave function $\varphi(t,\vx)$:
\be
	i \partial_t \varphi(t,\vx) = \big( H^\free + V(t,\vx) \big) \varphi(t,\vx),
	\label{eq:singlepartschroed}
\ee
where $H^\free$ denotes the free Hamiltonian, for example $H^\free = -\frac{1}{2m}\Delta$ or $H^\free = H^{\rm Dirac}$.\\
Let $G^\ret(t,\vx)$ be the retarded Green's function of the free  Schr{\"o}dinger equation, i.e.
\be
	\big( i \partial_t - H^\free \big) G^\ret(t,\vx) = \delta(t) \delta^{(3)}(\vx)
	\label{eq:schroedgreensfn}
\ee
with $G^\ret(t,\vx) = 0$ for $t< t_0$.
Then the initial value problem consisting of \eqref{eq:singlepartschroed} for $t\geq t_0$ with $\varphi(t_0,\vx) = \varphi_0(\vx)$ is equivalent to the integral equation (see \cite{feynman_positrons})
\be
	\varphi(t,\vx) = \varphi^\free(t,\vx) + \int_{t_0}^\infty \! dt' \int d^3 \vx' \, G^\ret(t-t',\vx-\vx') V(t',\vx') \varphi(t',\vx'),
	\label{eq:singlepartint}
\ee
where $\varphi^\free(t,\vx) = [e^{-i H^\free (t-t_0)} \varphi_0](\vx)$ is the respective solution of the free  Schr{\"o}dinger equation. The time integral in \eqref{eq:singlepartint} effectively only runs up to $t$ as $G^\ret(t-t',\vx-\vx') = 0$ for $t < t'$. One can easily check that a solution of \eqref{eq:singlepartint} satisfies \eqref{eq:singlepartschroed} by acting with $(i \partial_t - H^\free)$ on both sides of \eqref{eq:singlepartint} and using the Green's function property \eqref{eq:schroedgreensfn}. The initial conditions are satisfied as for $t=t_0$ the integral vanishes and $\varphi(t_0,\vx) = \varphi^\free(t_0,\vx) = \varphi_0(\vx)$. Conversely, we obtain \eqref{eq:singlepartint} from \eqref{eq:singlepartschroed} by inverting the operator $(i \partial_t - H^\free)$.
Note that the reformulation is exact (non-perturbative).

For two non-relativistic particles, one can proceed analogously.\footnote{In fact, this case is mathematically equivalent to the case of a single particle in 6 space dimensions.} The initial value problem $\varphi(t_0,\vx_1,\vx_2) = \varphi_0(\vx_1,\vx_2)$ of the  Schr{\"o}dinger equation
\be
	i \partial_t \varphi(t,\vx_1,\vx_2) = \big(H_1^\free + H_2^\free + V(t,\vx_1,\vx_2)\big) \varphi(t,\vx_1,\vx_2)
	\label{eq:twopartschroed}
\ee
for $t \geq t_0$ is equivalent to the integral equation
\begin{align}
	\varphi(t,\vx_1,\vx_2) = \varphi^\free(t,\vx_1,\vx_2) + \int_{t_0}^\infty \! dt' \! \int d^3 \vx_1'\,  d^3 \vx_2'\, &G_1^\ret(t-t',\vx_1-\vx_1') G_2^\ret(t-t',\vx_2-\vx_2')\nonumber\\
	\times&V(t',\vx_1',\vx_2') \varphi(t',\vx_1',\vx_2'),
	\label{eq:twopartschroedint}
\end{align}
where $\varphi^\free(t,\vx_1,\vx_2) = [e^{-i(H_1^\free + H_2^\free)(t-t_0)}\varphi_0](\vx_1,\vx_2)$ and $G_k^\ret(t,\vx_k)$ are the retarded Green's functions of the operators $(i\partial_t - H_k^\free)$. Note that the synchronized product $G_1^\ret(t,\vx_1) G_2^\ret(t,\vx_2)$ is the retarded Green's function of the operator $(i\partial_t - H_1^\free - H_2^\free)$.

While \eqref{eq:twopartschroedint} constitutes a viable equation for non-relativistic quantum physics, it is clearly not Lorentz invariant as it makes use of just a single time variable. Let us therefore pause and reconsider the step from the single-particle to the two-particle integral equation.\\
As we aim at a manifestly Poincar\'{e} invariant equation, it is helpful rewrite the single-particle integral equation \eqref{eq:singlepartint} in terms of spacetime variables:
\be
	\psi(x) = \psi^\free(x) + \int d^4 x' \, G^\ret(x-x') V(x') \psi(x'),
	\label{eq:singlepartintspacetime}
\ee
where we wrote $\psi$ instead of $\varphi$. In order to obtain a Poincar\'{e} invariant domain of integration, we take the initial time $t_0$ to $-\infty$. Note that the limits of the time integration as well as the choice of the Green's function mainly play a role for the initial value problem. A solution of the integral equation solves  Schr{\"o}dinger's equation regardless of these choices.

Now \eqref{eq:singlepartintspacetime} straightforwardly suggests a two-particle generalization in terms of a multi-time wave function:
\be
	\psi(x_1,x_2) = \psi^\free(x_1,x_2) + \int d^4 x_1'\, d^4 x_2' \, G_1(x_1-x_1') G_2(x_2-x_2') K(x_1',x_2') \psi(x_1',x_2').
	\label{eq:twopartintgeneral}
\ee
Here, the integration runs over $\M \times \M$ and $\psi^\free$ is (similarly to $\varphi^\free$ before) a given solution of the free multi-time equations. These are assumed to be covariant equations of the form
\be
	D_k  \psi^\free(x_1,x_2) = 0,~~k=1,2,
	\label{eq:freemultitime}
\ee
where $D_k$ is a first-order covariant differential operator, the main example being the Dirac operator $D_k = ( -i \gamma_k^\mu \partial_{k,\mu} + m_k)$.
Finally, $G_k(x_k)$ is a Green's function of the $k$-th free multi-time equation, i.e.
\be
	D_k G_k(x_k) = \delta^{(4)}(x_k).
\ee
In the following, we use the notation $G$ for a generic Green's functions of some covariant operator. If we want to be more specific which case we are considering, we indicate this by introducing new symbols or indices.
Apart from retarded Green's functions, which are expected to be relevant to the past initial value problem, we also allow for different choices such as advanced, time-symmetric and Feynman Green's functions here. Note that \eqref{eq:twopartintgeneral} is only manifestly time reversal invariant if the Green's functions are.

$K(x_1,x_2)$ is an, as yet, arbitrary function (or distribution) which we call the \textit{interaction kernel}. $K$ deserves this name as it occurs in the same place in \eqref{eq:twopartintgeneral} as the interaction potential in \eqref{eq:twopartschroedint}. As we aim at a Poincar\'{e} invariant equation, $K$ must be a Poincar\'{e} invariant function of $x_1,x_2$. Perhaps the most natural possibility is that $K$ depends on $x_1,x_2$ only through the Minkowski distance $s^2(x_1,x_2) = (x_1-x_2)^\mu (x_1-x_2)_\mu$. Stipulating that interaction happens exactly at light-like separation, we are led to
\be
	K(x_1,x_2) = \kappa \; \delta(s^2(x_1,x_2)),
	\label{eq:twopartkernel}
\ee
where $\kappa$ is a Poincar\'{e} invariant prefactor that may include coupling constants such as charges $e_1,e_2$ and $\gamma$-matrices. If $\gamma$-matrices are admitted, $\kappa$ can be non-constant, as then Poincar\'{e} invariant quantities besides $s^2(x_1,x_2)$ exist. For example, $\kappa$ could depend on $\gamma_i^\mu (x_1-x_2)_\mu$ for $i=1,2$. (We come back to this possibility later, see Eq. \eqref{eq:modifiedkappa}.)

Thus, we arrive at the following class of integral equations:
\be
	\psi(x_1,x_2) = \psi^\free(x_1,x_2) + \int d^4 x_1' \, d^4 x_2' \; G_1(x_1-x_1') G_2(x_2-x_2') \, \kappa \, \delta(s^2(x_1',x_2')) \psi(x_1',x_2').
	\label{eq:twopartint}
\ee
The equation becomes fully specified by the choices of (a) the free multi-time equations \eqref{eq:freemultitime}, (b) the particular Green's functions of these equations, and (c) the prefactor $\kappa$.

\paragraph{Examples.}
\begin{enumerate}
	\item \textit{Dirac particles.} In this case, the Green's functions are commonly denoted by $S^\ret$, $S^\sym$, $S_F$ etc. In view of time-reversal invariance, the most natural choice is the symmetric Green's function  $S^\sym(x) = \frac{1}{2}( S^\ret(x)+S^\adv(x)$. Also the Feynman propagator $S_F$ is time-symmetric (see Appendix \ref{sec:greensfns} for details). A natural choice for $\kappa$ turns out to be:
	\be
		\kappa = i\, \lambda \, \gamma_1^\mu \gamma_{2,\mu},
		\label{eq:naturalkappa}
	\ee
	where $\lambda$ is a positive constant. Later, we shall see $\lambda = \frac{e_1e_2}{4\pi}$ where $e_1, e_2$ are the charges of the particles. The factor $i$ is necessary to obtain the correct non-relativistic limit (see Sec. \ref{sec:nonrellimit}). Then \eqref{eq:twopartint} becomes (we here admit generic Green's functions):
	\be
		\!\!\!\!\!\!\!\!\!\! \psi(x_1,x_2) = \psi^\free(x_1,x_2) +i\lambda \! \int \! d^4 x_1' \, d^4 x_2' \, S_1(x_1-x_1') S_2(x_2-x_2') \gamma_1^\mu \gamma_{2,\mu} \delta(s^2(x_1',x_2')) \psi(x_1',x_2').
		\label{eq:twopartintspin}
	\ee
	\item \textit{KG particles.} Even though we have motivated \eqref{eq:twopartint} for cases where the free multi-time equations are of first order, it is possible to consider \eqref{eq:twopartint} also for the case that free multi-time equations are KG equations (which are of second order). In this case, $\kappa \in \R$ and one uses the Green's functions of the KG equation (see Appendix \ref{sec:greensfns}) instead of $G_1,G_2$ .
\end{enumerate}

\subsection{Important features of the integral equation} \label{sec:features}

There are several points of interest about eq. \eqref{eq:twopartint}.
\begin{enumerate}
	\item  It is a manifestly Lorentz invariant equation. Furthermore, it is also manifestly time reversal invariant, provided one chooses time-symmetric Green's functions.
	\item Essential use is made of a multi-time wave function. Contrary to the differential multi-time equations \eqref{eq:multitimeeqs}, there is no frame in which \eqref{eq:twopartint} reduces to an autonomous equation in the single-time wave function $\varphi$ associated with that frame via \eqref{eq:singlemulti}. In this sense, the multiple times are indispensable for the dynamics.
	\item The derivation suggests that the role of eq. \eqref{eq:twopartint} may be to define the future time evolution of an initial free wave function $\psi^\free$ at large negative times $t_0$, at least for retarded Green's functions. However, we genuinely changed the structure of the non-relativistic integral equation \eqref{eq:twopartschroedint} to which this role applies (for $t_0 \rightarrow \infty$). So this role is not obvious anymore, in particular in the time-symmetric case. It shall be discussed in Sec. \ref{sec:evollaw} how \eqref{eq:twopartint} defines the time evolution of $\psi$.
	\item \textit{Non-relativistic limit.} As still unexplored equations, it is important to see how \eqref{eq:twopartint} relates to known equations in a suitable limit. In our case, the expected non-relativistic limiting equation is the  Schr{\"o}dinger equation. We shall demonstrate in Sec. \ref{sec:nonrellimit} that we do, indeed, obtain a  Schr{\"o}dinger equation with a Coulomb-type potential for the single-time wave function $\varphi$ if time delay effects in \eqref{eq:twopartint} are neglected.
	\item \textit{\eqref{eq:twopartint} as a boundary integral equation.} Note that the integral term in \eqref{eq:twopartint} only involves $\psi$ on light-like configurations, i.e., configurations on the boundary $\partial \mathscr{S}$ of $\mathscr{S}$. This implies that one can solve \eqref{eq:twopartint} on $\partial \mathscr{S}$ autonomously as a \textit{boundary integral equation}. Then \eqref{eq:twopartint} can be used as a formula to calculate $\psi$ in the interior of $\mathscr{S}$.
	Moreover, \eqref{eq:twopartint} can be rewritten as follows.	 Let
	\be
		d \sigma(x_1,x_2) = \delta(s^2(x_1,x_2)) \, d^4 x_1 \, d^4 x_2
		\label{eq:limeasure}
	\ee
	be the Poincar\'{e} invariant surface measure on $\partial \mathscr{S}$. Then \eqref{eq:twopartint} becomes:
	\be
		\psi(x_1,x_2) = \psi^\free(x_1,x_2) +  \int_{\partial \spacelike} \!\!\! d \sigma(x_1',x_2') \, G_1(x_1-x_1') G_2(x_2-x_2') \, \kappa \, \psi(x_1',x_2').
		\label{eq:twopartboundaryint}
	\ee
	It is striking that the interaction kernel has disappeared from the equation. Interaction effects only result from the use of the natural measure \eqref{eq:limeasure} on $\partial \spacelike$. This interesting observation can also be used to guess a particular $N$-particle integral equation (see Sec. \ref{sec:npartint}).
\end{enumerate}

\paragraph{Generalizations.}
Apart from the $N$-particle case which deserves a separate discussion (Sec. \ref{sec:npart}), we can immediately generalize \eqref{eq:twopartint} in the following regards.
\begin{enumerate}
	\item For \textit{different spacetime dimensions,} say $1+d$ with $d \in \N$, two ways of formulating \eqref{eq:twopartint} are conceivable. In both cases, we take $G_1, G_2$ to be Green's functions of the respective free quantum mechanical wave equations (e.g. the Dirac equation) associated with that spacetime dimension. For the interaction kernel, there are two possibilities: (a) Note that $\frac{1}{4\pi}\delta(s^2(x_1,x_2))$ is the time-symmetric causal Green's function of the wave equation. In $1+d$ spacetime dimensions, we can accordingly replace $\frac{1}{4\pi}\delta(s^2(x_1,x_2))$ by the time-symmetric causal Green's function of the respective wave equation. We list the Green's functions for $d=1,2,3$ in Appendix \ref{sec:greensfns}.
	(b) The second possibility is to simply use $\delta(s^2(x_1,x_2))$ for all spacetime dimensions.
	\item \textit{Curved spacetime.} Similarly to 1., there are the following two possibilities to extend the integral equation to curved spacetimes. In place of $G_i(x_i-x_i')$ one uses Green's functions $G_i(x_i,x_i')$ of the respective free wave equations on curved spacetime. Then possibility (a) corresponds to using instead of $\frac{1}{4\pi}\delta(s^2(x_1,x_2))$ the time-symmetric causal Green's function of the massless conformally invariant scalar wave equation (see \cite{john}) and, for \eqref{eq:naturalkappa}, the respective Dirac $\gamma$-matrices on curved spacetime (see \cite{pollock}). Possibility (b) corresponds to using $\delta(s^2(x_1,x_2))$ with $s^2(x_1,x_2)$ given by Synge's world function, i.e., the generalization of the Minkowski distance for curved spacetimes (see \cite[sec. 3]{Poisson_2011}).
\end{enumerate}

\subsection{Comparison with the Bethe-Salpeter equation} \label{sec:bs}
	The integral equation \eqref{eq:twopartint} is similar to the so-called the "ladder approximation" of the Bethe-Salpeter (BS) equation \cite{wick_54,nakanishi,greiner_qed}. The BS equation contains an infinite sum over Feynman diagrams in its interaction kernel and is more a framework, not \textit{per se} a well-defined closed equation.  In the ladder approximation, one only takes into account a subclass of diagrams, namely those which correspond to the "exchange of only one virtual photon at a time" \cite[p. 332]{greiner_qed}. By this procedure, one obtains a closed Poincar\'{e} invariant equation of the form \eqref{eq:twopartintgeneral}.
	 Compared to \eqref{eq:twopartint}, the ladder approximation of the BS equation (BSL) is fixed on the choice of Feynman propagators $G_1^F, G_2^F$. $K(x_1,x_2)$ then corresponds to the Feynman propagator of the respective exchange particle in the QFT picture, for example for photons:
	 \be
	 	K(x_1,x_2) = i \, e^2 \gamma_1^\mu \gamma_{2,\mu} \, D_F(x_1-x_2).
	 	\label{eq:bslkernel}
	 \ee 
	Here, $D_F(x)$ stands for the Feynman propagator of massless scalar particles.

	Compared to the BSL equation, \eqref{eq:twopartint} has several differences, both (a) in its concrete mathematical form, as well as (b) in its relation to other theories. This, in turn, has different implications for the generalizations we shall discuss in the later sections of this paper. With regard to (a), the mathematical form of the equation, we note the following points.
\begin{enumerate}
	\item[(i)] \eqref{eq:twopartint} admits the choices of the symmetric and retarded Green's functions while BSL does not. Thus, \eqref{eq:twopartint} defines a wider class of equations than BSL.
	\item[(ii)] As we discussed, the natural choice of $K(x_1,x_2)$ in \eqref{eq:twopartint} is $K(x_1,x_2) = \kappa \, \delta(s^2(x_1,x_2))$. This is similar to \eqref{eq:bslkernel}, with $\kappa =i \frac{e^2}{4\pi} \gamma_1^\mu \gamma_{2,\mu}$ and $\delta(s^2(x_1,x_2))$ instead of $D_F(x_1-x_2)$. (In \eqref{eq:naturalkappa} we then have $\lambda=\frac{e^2}{4\pi} \approx 1/137$.) The function $\frac{1}{4\pi} \delta(s^2(x_1,x_2))$ is, like $D_F(x_1-x_2)$, a Green's function of the massless Klein-Gordon equation. However, while $D_F(x_1-x_2)$ has support also on time-like and space-like configurations, $\delta(s^2(x_1,x_2))$ only has support on light-like configurations. As a consequence, \eqref{eq:twopartint} expresses interaction exactly along light cones while BSL does not. 
Furthermore, this implies that \eqref{eq:twopartint} is an equation on the natural domain $\overline{\spacelike}$ of multi-time wave functions. By contrast, BSL necessarily involves $\psi(x_1,x_2)$ on the whole of $\M\times \M$. 
\end{enumerate}

	This leads us to (b), the relation with other theories. The BSL equation is derived from QFT while \eqref{eq:twopartint} was motivated by an independent heuristic consideration. It is therefore not surprising that the multi-time wave functions in the two equations have a different meaning. The BS wave function $\psi_{\rm BS}$ is usually defined via QFT by
	\be
		\psi_{\rm BS}(x_1,x_2) = \langle 0 | \widehat{T}( \widehat{\psi}(x_1) \widehat{\psi}(x_2)) | \psi_H \rangle,
		\label{eq:bsheisenberg}
	\ee
	where $\widehat{\psi}(x)$ denotes the Heisenberg field operator, $\widehat{T}$ the time-ordering symbol and $| \psi_H \rangle$ the Heisenberg state vector.

	By contrast, a multi-time wave function need not be defined with appeal to QFT. Only in certain cases (essentially for local quantum field theories), does the following relation hold on space-like configurations $(x_1,x_2) \in \spacelike$ \cite{qftmultitime}:
	\be
		\psi(x_1,x_2) = \langle 0 | \widehat{\psi}(x_1) \widehat{\psi}(x_2) | \psi_H \rangle.
		\label{eq:multitimeheisenberg}
	\ee	
	The difference between \eqref{eq:bsheisenberg} and \eqref{eq:multitimeheisenberg} shows that one should not, in general, expect the BS wave function and the multi-time wave function to coincide. What is more, if we do not start from a QFT in the Heisenberg picture there need not be any expression of $\psi(x_1,x_2)$ via Heisenberg field operators.

	These points show that the integral equation \eqref{eq:twopartint} and the BSL equation are different equations, despite their structural similarity. Accordingly, starting from \eqref{eq:twopartint} or \eqref{eq:integrodiff}, different generalizations to the $N$-particle case appear natural. This creates interesting new possibilities which shall be explored in Sec. \ref{sec:npart}, after more basic questions have been clarified.

\subsection{The integral equation \eqref{eq:twopartint} as an evolution law} \label{sec:evollaw}

We would like to regard \eqref{eq:twopartint} as an evolution law, i.e., a law that determines $\psi$ uniquely for all times, given certain data. As \eqref{eq:twopartint} contains integrals over all future and past configurations (in the time-symmetric case) or at least all past configurations (in the retarded case), it does not admit a Cauchy initial value problem. However, we can say more about the role of \eqref{eq:twopartint} by considering its abstract structure as an operator equation on a suitable Banach space $\Banach$. Let $\hat{K}$ denote the integral operator in \eqref{eq:twopartint}. Then \eqref{eq:twopartint} has the form
\be
	\psi = \psi^\free + \widehat{K} \psi.
	\label{eq:operatoreq}
\ee
 There are two scenarios where the question of the existence and uniqueness of solutions can be answered in a straightforward way.
\begin{enumerate}
	\item[(a)] If $\widehat{K}$ is a contraction, i.e., $\| \widehat{K} \| < 1$, then Banach's fixed point theorem ensures the existence of a unique solution for every $\psi^\free \in \Banach$.
	\item[(b)] If $\widehat{K}$ is a compact operator on $\Banach$, then the Fredholm alternative applies \cite{kress}. This means: if the solutions of \eqref{eq:operatoreq} are unique then there also \textit{exists} a solution for every $\psi^\free \in \Banach$. The uniqueness of solutions has to be shown by a separate method; assume, for the moment, that such a method has been found.
\end{enumerate}
The crucial point now is that for both (a) and (b) the free solution $\psi^\free$ determines the solution $\psi$ uniquely, i.e., $\psi^\free$ plays the role of \textit{determining data}. This is not the same as \textit{initial data}, but more can be said about the relation between these two notions.
\begin{enumerate}
	\item[(i)] The function $\psi^\free(t_1,\vx_1,t_2,\vx_2)$ (determining data) is, as a free solution of the multi-time equations \eqref{eq:freemultitime}, itself determined uniquely by Cauchy data, e.g., $\psi^\free(t_0,\vx_1,t_0,\vx_2)=\psi_0(\vx_1,\vx_2)$, for some $t_0 \in \R$. That means, the determining data of the integral equation \eqref{eq:twopartint} are Cauchy data $\psi_0$ for $\psi^\free$.
	\item[(ii)] However, $\psi_0$ does not play the role of initial data. That is because in general $\psi(t_0,\vx_1,t_0,\vx_2) \neq \psi^\free(t_0,\vx_1,t_0,\vx_2)$. Nevertheless, we can read \eqref{eq:twopartintgeneral} as determining a correction to the free solution as a consequence of interaction.
	\item[(iii)] In the case of retarded Green's functions one can say more, at least on a heuristic level. Assume the integral in \eqref{eq:twopartint} does not contribute much for large negative times $t_1,t_2$. Then we find that for $t_1,t_2 \rightarrow -\infty$, $\psi(t_1,\cdot,t_2,\cdot)$ approaches $\psi^\free(t_1,\cdot,t_2,\cdot)$. In that case, we arrive at a \textit{scattering picture:} \eqref{eq:twopartint} determines $\psi$ given the free incoming wave function $\psi^\free$ for $t_1,t_2 \rightarrow -\infty$.
\end{enumerate}
We conclude that in the cases (a) and (b), the multi-time integral equation \eqref{eq:twopartint} indeed has the role of an evolution law. In view of the restrictive consistency conditions \eqref{eq:cc} for differential multi-time equations \eqref{eq:multitimeeqs} which make interacting dynamics difficult to achieve, this is striking and deserves further investigation. In the following section we shall reformulate the integral equation \eqref{eq:twopartint} as a system of two integro-differential equations which have a similar structure as the previous multi-time equations \eqref{eq:multitimeeqs}. This allows for a comparison between these two types of multi-time equations, in particular with regard to the consistency question. Moreover, the integro-differential equations facilitate the discussion of the non-relativistic limit (Sec. \ref{sec:nonrellimit}) and are easier to interpret physically (Sec. \ref{sec:wf}).

\subsection{Multi-time integro-differential equations} \label{sec:integrodiff}

Acting on both sides of \eqref{eq:twopartint} with one of the free operators $D_k$ \eqref{eq:freemultitime} for each $k=1,2$ separately and then collecting both resulting equations in one system yields:\footnote{Note that this way of deriving a system of multi-time equations from \eqref{eq:twopartint} is analogous to the way one obtains the usual  Schr{\"o}dinger equation from \eqref{eq:singlepartint}.}
\begin{align}
	 	(D_1 \psi)(x_1,x_2) &= \int d^4 x_2' ~ G_2(x_2-x_2') \, \kappa \, \delta(s^2(x_1,x_2')) \, \psi(x_1,x_2'), \nonumber\\
	 	(D_2 \psi)(x_1,x_2) &= \int d^4 x_1' ~ G_1(x_1-x_1') \, \kappa \, \delta(s^2(x_1',x_2)) \, \psi(x_1',x_2),
	 	\label{eq:integrodiff}
\end{align}
where $\psi^\free$ has dropped out as $D_k \psi^\free = 0$ for $k=1,2$.\\
Thus, every solution of \eqref{eq:twopartint} satisfies \eqref{eq:integrodiff}. Conversely, let $\psi$ be a solution of \eqref{eq:integrodiff}. Then inverting the operators $D_k$ yields the two equations:
\be
	 \psi(x_1,x_2) = \psi^\free_k(x_1,x_2) + \int d^4 x_1' \, d^4 x_2' \, G_1(x_1-x_1') G_2(x_2-x_2') \, \kappa \, \delta(s^2(x_1',x_2')) \, \psi(x_1',x_2'),
\ee
for $k=1,2$ where $\psi^\free_k$ is a solution $D_k \psi^\free_k = 0$. Taking the difference of the two equations implies $\psi_1^\free = \psi_2^\free$. So the two equations coincide and we find that $\psi$ satisfies \eqref{eq:twopartint} for a certain $\psi^\free$. Therefore, \eqref{eq:twopartint} and \eqref{eq:integrodiff} are equivalent.

\eqref{eq:integrodiff} is a system of $N=2$ multi-time equations similar to the one discussed earlier in Sec. \ref{sec:multitime} (Eq. \eqref{eq:multitimeeqs}). However, the interaction terms on the right hand side of \eqref{eq:integrodiff} are integral operators whereas in \eqref{eq:multitimeeqs} only differential operators are admitted. We are thus led to consider more general types of multi-time equations,
\be
	(D_k \psi)(x_1,x_2) = (L_k \psi)(x_1,x_2),
	\label{eq:generalmultitime}
\ee
for $k=1,2$ where the $L_k$ are integral operators.

In general, one would not expect these equations to have any non-trivial solutions at all. However, we saw already in the previous section that in certain scenarios \eqref{eq:twopartint} possesses a wide class of solutions. As the system \eqref{eq:integrodiff} is equivalent to \eqref{eq:twopartint}, it has solutions in theses cases as well. Thus, \eqref{eq:integrodiff} is special in the class \eqref{eq:generalmultitime}: the fact that it derives from a single integral equation creates a certain harmony of the interaction terms which allows \eqref{eq:generalmultitime} to have solutions. 

By contrast, inverting the operators $D_1,D_2$ in the general system of multi-time equations leads to the two equations
\begin{align}
	\psi(x_1,x_2) = \psi^\free_1(x_1,x_2) + \int d^4 x_1' \, G_1(x_1-x_1') (L_1 \psi)(x_1',x_2),\nonumber\\
	\psi(x_1,x_2) = \psi^\free_2(x_1,x_2) + \int d^4 x_2' \, G_2(x_2-x_2') (L_2 \psi)(x_1,x_2'),
	\label{eq:generalmultitimeinverted}
\end{align}
where $\psi^\free_k,~k=1,2$ are again free solutions of $D_k \psi^\free_k = 0$. Now the crucial point is: unless $L_1, L_2$ have a special relation, these two equations are different and it is unclear whether they have any non-trivial solutions. Thus for multi-time equations \eqref{eq:generalmultitime} which can be recast as a single integral equation, the question of the existence of solutions is less problematic.

\paragraph{On the notion of consistency of \eqref{eq:integrodiff}.} The above discussion shows that there is a sense in which the multi-time system \eqref{eq:integrodiff} is consistent: the many multi-time equations do not conflict with each other (they derive from a single integral equation) and the system has a wide class of solutions. We now compare this sense with the notion of consistency for differential multi-time equations. For these, the role of the consistency condition \eqref{eq:cc} is to ensure the existence of solutions \textit{for arbitrary Cauchy data.} For \eqref{eq:integrodiff} by contrast, there is a solution $\psi$ for every solution $\psi^\free$ of the free multi-time equations. As Cauchy data are in a one-to-one correspondence with free solutions, \eqref{eq:integrodiff} possesses as many solutions as a system of differential multi-time equations does. Nevertheless, it is not clear whether for \textit{arbitrary Cauchy} data there is a solution $\psi$ compatible with these data. The sense in which \eqref{eq:integrodiff} is consistent is therefore weaker than the sense of consistency for differential multi-time equations. However, to insist on solutions for arbitrary Cauchy data also for \eqref{eq:integrodiff} would be unnatural because of the time delay.

\paragraph{Consistency of multi-time integro-differential equations in the retarded case.} The previous paragraph shows that the notion of the consistency of multi-time equations is relative to the data which classify the solutions. In case the interaction terms $L_k \psi$ in \eqref{eq:generalmultitime} depend only on the past, a new notion of data and solutions becomes available: specifying an entire history $\chi(t_1,\vx_1,t_2,\vx_2)$ up to times $t_1,t_2 \leq t_0$ for some $t_0$ and then looking for a function $\psi(t_1,\vx_1,t_2,\vx_2)$ of \eqref{eq:generalmultitime} that agrees with this history for $t_1,t_2 \leq t_0$ and satisfies the multi-time equations for $t_1,t_2 > t_0$. We then call $\psi$ a \textit{conditional solution} with history $\chi$. Demanding that conditional solutions exist for arbitrary histories leads to a new consistency condition that is similar in spirit to the consistency condition \eqref{eq:cc} for differential multi-time equations. We show this in detail in Appendix \ref{sec:ccret}. The upshot of the discussion is that the resulting consistency condition is automatically satisfied if the multi-time equations \eqref{eq:generalmultitime} derive from an integral equation of the form \eqref{eq:twopartintgeneral}.

\vspace{0.2cm}
While the integral equation \eqref{eq:twopartint} is thus advantageous for discussing the existence of solutions, its reformulation as the system of integro-differential equations \eqref{eq:integrodiff} also has its merits, in particular for physical considerations. To illustrate this, we now use \eqref{eq:integrodiff} to study the non-relativistic limit. Moreover, in Sec. \ref{sec:wf}, we show that the interaction terms in \eqref{eq:integrodiff}  have strong analogies with WF electrodynamics, and are therefore easier to interpret than \eqref{eq:twopartint}.

\subsection{Non-relativistic limit} \label{sec:nonrellimit}

The integral equation \eqref{eq:twopartint} has been motivated starting from a single-time  Schr{\"o}dinger equation. However, significant modifications were made during its derivation such that a separate consideration becomes necessary to see how one can obtain a  Schr{\"o}dinger equation for the single-time wave function $\varphi(t,\vx_1,\vx_2) = \psi(t,\vx_1,t,\vx_2)$ associated with a particular frame in suitable limiting cases. Such a consideration shall be presented here.\footnote{A first version of the argument was given in \cite[p. 135]{phd_lienert}. I am grateful to Lukas Nickel for a discussion about the case with spin.} We focus on the Dirac case (Eq. \eqref{eq:twopartintspin}) and use retarded Green's functions.

The derivation rests on one crucial assumption, namely that \textit{time delay is negligible} in integrals over $\psi$. We shall use this assumption by replacing expressions such as $\psi(t,\vx_1,t \pm |\vx_1-\vx_2|,\vx_2)$ with $\psi(t,\vx_1,t,\vx_2)$.
This replacement is clearly adequate only for certain wave functions, namely those which vary only relatively slowly in time intervals corresponding to (in natural units) the spatial distances $|\vx_1-\vx_2|$ that occur in the support of $\psi$.

The usual way to derive single-time equations from multi-time equations is to consider the time derivative $i \partial_t \varphi(t,\vx_1,\vx_2) = i \partial_t \psi(t,\vx_1,t,\vx_2)$. Then one uses the chain rule,
\be
	i \partial_t \varphi(t,\vx_1,\vx_2) = [i \partial_{t_1} \psi(t_1,\vx_1,t_2,\vx_2) + i \partial_{t_2} \psi(t_1,\vx_1,t_2,\vx_2)]|_{t_1=t_2=t},
	\label{eq:chainrule}
\ee
and expresses the time derivatives $i \partial_{t_k}\psi$ using the multi-time equations. Using the multi-time integro-differential equations \eqref{eq:integrodiff} of the previous section, which are equivalent to \eqref{eq:twopartint}, allows us to make an analogous consideration here.

We start by expressing $i \partial_{t_k} \psi$ in \eqref{eq:chainrule} via \eqref{eq:integrodiff} with $G_i = S^\ret_i$ (the retarded Dirac propagators)\footnote{More generally, the derivation leads to the same result if the identity $S_k(0,\vx_k) = i \gamma^0_k \delta^{(3)}(\vx_k)$ holds, a property of many Green's functions.}. In order to be able to use  \eqref{eq:integrodiff}, we multiply the first equation in \eqref{eq:integrodiff} with $\gamma^0_1$ and the second with $\gamma_2^0$, noting that $\gamma_k^0 D_k = (-i \partial_{t_k} + H_k^{\rm Dirac})$. We find:
\be
	i \partial_t \psi(t,\vx_1,t,\vx_2) = \big( H_1^{\rm Dirac} + H_2^{\rm Dirac} \big) \psi(t,\vx_1,t,\vx_2) - I_1(t,\vx_1,\vx_2) - I_2(t,\vx_1,\vx_2),
	\label{eq:almostschroed}
\ee
where the interaction terms $I_1, I_2$ are given by
\begin{align}
	I_1(t,\vx_1,\vx_2) &= \gamma_1^0\int d t_2' \,d^3 \vx_2' ~  S_2^\ret(t-t_2',\vx_2-\vx_2') \, \kappa \, \delta \big( s^2((t,\vx_1),(t_2',\vx_2')) \big) \, \psi(t,\vx_1,t_2',\vx_2'), \nonumber\\
	I_2(t,\vx_1,\vx_2) &= \gamma_2^0 \int d t_1' \, d^3 \vx_1' ~ S_1^\ret(t-t_1',\vx_1-\vx_1') \, \kappa \, \delta \big(s^2((t_1',\vx_1'),(t,\vx_2)) \big) \,\psi(t_1',\vx_1,t,\vx_2').
\end{align}
Consider $I_1$. We decompose $\delta(s^2((t_1,\vx_1),(t_2,\vx_2))) = \delta((t_1-t_2)^2 - |\vx_1-\vx_2|^2)$ as
\be
	\delta((t_1-t_2)^2 - |\vx_1-\vx_2|^2) = \frac{1}{2 |\vx_1-\vx_2|} \big[ \delta(t_1-t_2 - |\vx_1-\vx_2|) + \delta(t_1-t_2 + |\vx_1-\vx_2|) \big].
	\label{eq:deltadecomp}
\ee
Using \eqref{eq:deltadecomp} in the expression for $I_1$ and performing the $t_2'$-integration yields:
\begin{align}
	I_1 = \gamma_1^0 \int d^3 \vx_2' ~  &\left[ S_2^\ret(\varepsilon,\vx_2-\vx_2') \, \frac{\kappa}{2 |\vx_1-\vx_2'|} \psi(t,\vx_1,t - \varepsilon,\vx_2') \right. \nonumber\\
& \left. + S_2^\ret(-\varepsilon,\vx_2-\vx_2') \, \frac{\kappa}{2 |\vx_1-\vx_2'|} \psi(t,\vx_1,t + \varepsilon,\vx_2') \right],
\end{align}
where $\varepsilon = |\vx_1-\vx_2'|$. The term in the second line vanishes as $S^\ret_i(t,\vx) = 0$ for $t<0$. We now neglect the time delay. Then:
\be
	I_1 \approx \int d^3 \vx_2' ~ \lim_{\varepsilon \rightarrow 0} S_2^\ret(\varepsilon,\vx_2-\vx_2')\, \frac{\kappa}{2 |\vx_1-\vx_2'|} \psi(t,\vx_1,t,\vx_2').
\ee
Next, we make use of the property $\lim_{\varepsilon \rightarrow 0} S_2^\ret(\varepsilon,\vx_2-\vx_2') = i \gamma_2^0 \delta^{(3)}(\vx_2-\vx_2')$ (see Appendix \ref{sec:greensfns}). This results in:
\be
	I_1 \approx \frac{i \,\gamma_1^0 \gamma_2^0 \, \kappa}{2|\vx_1-\vx_2|} \, \psi(t,\vx_1,t,\vx_2).
\ee
Together with an analogous consideration for $I_2$ we finally obtain (considering \eqref{eq:almostschroed}):
\be
	i \partial_t \varphi(t,\vx_1,\vx_2) \approx \left( H_1^{\rm Dirac} + H_2^{\rm Dirac} - \frac{i \,\gamma_1^0 \gamma_2^0 \, \kappa}{|\vx_1-\vx_2|} \right) \varphi(t,\vx_1,\vx_2).
	\label{eq:freeschroedcoulomb}
\ee
This is the usual single-time  Schr{\"o}dinger equation with a spin-dependent Coulomb potential.
For $\kappa = i \lambda \, \gamma_1^\mu \gamma_{2,\mu}$ \eqref{eq:naturalkappa}, we obtain:
\be
		V(t,\vx_1,\vx_2) =\lambda \left( \frac{1}{|\vx_1-\vx_2|} - \frac{\boldsymbol{\alpha}_1 \cdot \boldsymbol{\alpha}_2}{|\vx_1-\vx_2|} \right),
		\label{eq:coulombspin}
	\ee
where $\boldsymbol{\alpha}_k = (\gamma^0_k \gamma_k^j)$, $j=1,2,3$.

The comparison with the usual prefactor of the Coulomb potential now also allows us to determine the value of $\lambda$, namely
\be
	\lambda = \frac{e_1 e_2}{4\pi},
\ee
where $e_1,e_2$ are the charges of the two particles. For $e_1 = e_2 = e$ we can thus identify $\lambda$ with the fine structure constant $\alpha \approx 1/137$.

\paragraph{Remarks.}
\begin{enumerate}
	\item Our derivation appears to be shorter and more transparent than the analysis which is commonly used for the non-retarded limit of the BS equation (see e.g. \cite[pp. 344--350]{greiner_qed}).
	\item On the first glance it is, perhaps, surprising that the integro-differential equations \eqref{eq:integrodiff} (or equivalently the integral equation \eqref{eq:twopartint}), reduce to an autonomous differential equation under the given assumptions. This is a special feature of the interaction kernel $K(x_1,x_2) \propto \delta(s^2(x_1,x_2))$. Neglecting time delay, the light cones involved in $\delta(s^2(x_1,x_2))$ turn into simultaneity surfaces (in the given frame). The resulting synchronization leads to a single-time equation.
	\item From the viewpoint of the integral equation \eqref{eq:twopartint}, the above consideration gives a justification why a potential occurs in the single-time  Schr{\"o}dinger equation, and furthermore shows that this potential should be a Coulomb-type potential. The crucial input that leads to this conclusion is the geometrically natural choice of the interaction kernel $\kappa \, \delta(s^2(x_1,x_2))$.
	\item \textit{Comparison with the Breit potential.} The resulting potential \eqref{eq:coulombspin} is to be compared with the Breit potential \cite{breit_eq}, a semi-relativistic potential used in nuclear physics:
	\be
		V_{\rm Breit}(t,\vx_1,\vx_2)  = \lambda \left( \frac{1}{|\vx_1-\vx_2|} - \frac{\boldsymbol{\alpha}_1 \cdot \boldsymbol{\alpha}_2}{2|\vx_1-\vx_2|}  - \frac{(\boldsymbol{\alpha}_1 \cdot (\vx_1-\vx_2)) \, (\boldsymbol{\alpha}_2 \cdot (\vx_1-\vx_2))}{2|\vx_1-\vx_2|^3} \right).
		\label{eq:breit}
	\ee
	We see that the respective first terms of \eqref{eq:coulombspin} and \eqref{eq:breit} agree. The remaining terms are similar but not identical.\\
In order to exactly obtain the Breit potential, one may consider modifying the interaction kernel of \eqref{eq:twopartintspin}. Such a procedure is discussed in \cite[pp. 344--350]{greiner_qed} for the BSL equation but it involves giving up manifest Lorentz invariance. Here we note that there exist further manifestly covariant choices for the factor $\kappa$ if one admits it to depend on $x_1-x_2$. For example,
\be
	\kappa(x_1,x_2) = i \lambda \left( \gamma_1^\mu \gamma_{2,\mu}+ \frac{1}{2}\frac{\gamma_1^\nu (x_1-x_2)_\nu \, \gamma_2^\rho (x_1-x_2)_\rho}{s^2(x_1,x_2)} \right).
	\label{eq:modifiedkappa}
\ee
Repeating the previous derivation for this choice leads to the appearance of $\kappa(0,\vx_1,0,\vx_2)$ instead of the previous $\kappa$ in \eqref{eq:freeschroedcoulomb}, and altogether one obtains the interaction potential
\be
	V(t,\vx_1,\vx_2)  = \lambda \left( \frac{1}{|\vx_1-\vx_2|} - \frac{\boldsymbol{\alpha}_1 \cdot \boldsymbol{\alpha}_2}{|\vx_1-\vx_2|}  - \frac{(\boldsymbol{\alpha}_1 \cdot (\vx_1-\vx_2)) \, (\boldsymbol{\alpha}_2 \cdot (\vx_1-\vx_2))}{2|\vx_1-\vx_2|^3} \right).
		\label{eq:almostbreit}
\ee
This is almost identical to the Breit potential \eqref{eq:breit}; however, there is a factor of $\frac{1}{2}$ missing in the second term. As the first two terms in \eqref{eq:almostbreit} result from  the term $\gamma_1^\mu \gamma_{2,\mu}$ in $\kappa$, it is puzzling how one could obtain a relative factor of $\frac{1}{2}$ between these two terms starting from a manifestly Lorentz invariant equation.\footnote{Perhaps this factor could have something to do with negative energies. Neglecting solutions with negative energies seems to be a common practice in derivations of the Breit equation from the BS equation (see e.g. \cite[p. 349]{greiner_qed}).}\\
As we are concerned with more basic conceptual questions here, we content ourselves with a plausible potential for the moment and turn to the physical interpretation of the interaction terms in \eqref{eq:integrodiff}.
\item \textit{Yukawa potential.} In case one chooses a Green's function of the KG equation for the interaction kernel, i.e., $K(x_1,x_2) = 4\pi i \, \lambda \, G_{KG}(x_1-x_2)$, one obtains a Yukawa potential by making a similar approximation as before. Namely, if we neglect the $p^0$-dependence in the Fourier transform $\widetilde{G}_{\rm KG}(p)$ of $G_{\rm KG}(x)$, we obtain:
\be
	\widetilde{G}_{\rm KG}(p) = \frac{1}{-p^2 + m^2} ~ \rightarrow ~ \frac{1}{\mathbf{p}^2 + m^2}.
\ee
Now we transform back to position space using the identity
\be
	\int \frac{d^3 p}{(2\pi)^3} \frac{e^{i \mathbf{p} \cdot x}}{\mathbf{p}^2 + m^2} = \frac{1}{4\pi} \frac{e^{-m |\vx|}}{|\vx|}.
\ee
The $p^0$-integration yields a $\delta$-function in $x^0$. Overall, the approximation leads to
\be
	G_{KG}(x_1-x_2) ~\rightarrow ~ \frac{\delta(t_1-t_2)}{4\pi} \frac{e^{-m |\vx_1-\vx_2|}}{|\vx_1-\vx_2|}.
\label{eq:yukawaapprox}
\ee
One can see that for $m=0$, the approximation has the same effect as neglecting the time delay in $\delta(s^2(x_1,x_2))$. Therefore, we regard it as the analog of the latter in case of $m>0$. Using the same steps as in the previous derivation, we obtain:
\be
	V(t,\vx_1,\vx_2) = \lambda\, \gamma_1^0 \gamma_2^0 \,\frac{e^{-m |\vx_1-\vx_2|}}{|\vx_1-\vx_2|}.
\ee
This is the Yukawa potential (times $\gamma_1^0 \gamma_2^0$). We believe that by choosing Green's functions of appropriate relativistic wave equations in the interaction kernel, one can also reproduce other types of effective potentials that actually occur in applications.
\end{enumerate}

\subsection{Action-at-a-distance form of the multi-time integro-differential equations} \label{sec:wf}

The fact that the interaction terms in the integro-differential multi-time equations \eqref{eq:integrodiff} reduce to a Coulomb-type potential when time delay is neglected shows that \eqref{eq:integrodiff} is more directly related to the known physical framework than the integral equation \eqref{eq:twopartint}. In the following we analyze the structure of \eqref{eq:integrodiff} also for situation where time delay is not negligible. This allows us to draw parallels with the Wheeler-Feynman formulation of classical electrodynamics.

Note that the multi-time system \eqref{eq:integrodiff} for Dirac particles can be rewritten as follows (recall $D_k = (-i \gamma_k^\mu \partial_{k,\mu}+ m_k)$):
\begin{equation}
	[ -i \gamma_k^\mu \partial_{k,\mu}- e_k \gamma_k^\mu \widehat{A}_{3-k,\mu}(x_k) + m_k ]\psi(x_1,x_2)=0,~~k=1,2.
 \label{eq:integrodiff2}
\end{equation}
Each of these equations is a Dirac equation in the respective particle index. They contain minimally coupled "field terms", given by:
\begin{align}
   [\widehat{A}_{1,\mu}(y) \psi](x_1,x_2) &=  i\frac{e_1}{4\pi} \int d^4 x_1' \, S_1(x_1-x_1') \gamma_{1,\mu} \, \delta(s^2(y,x_1')) \, \psi(x_1',x_2), \nonumber\\
[\widehat{A}_{2,\mu}(y) \psi](x_1,x_2) &= i\frac{e_2}{4\pi}  \int d^4 x_2' \, S_2(x_2-x_2') \gamma_{2,\mu} \, \delta(s^2(y,x_2')) \, \psi(x_1,x_2').
\label{eq:fieldterms}
\end{align}
Here, we have used $\lambda = \frac{e_1 e_2}{4\pi}$. We note in passing that one would also obtain minimal coupling of the interaction terms if $\kappa = i \,\lambda \, \gamma_1^\mu \gamma_{2,\mu}$ in the integral equation \eqref{eq:twopartint} was replaced by $i \,\lambda \, \gamma_1^\mu \gamma_2^\nu \kappa_{\mu \nu}(x_1,x_2)$ for some Poincar\'{e} invariant $\kappa_{\mu \nu}(x_1,x_2)$ such as $\kappa_{\mu \nu}(x_1,x_2) = (x_1-x_2)_\mu (x_1-x_2)_\nu /s^2(x_1,x_2)$, as in \eqref{eq:modifiedkappa}. That being said, we continue with $\kappa=i\, \lambda \, \gamma_1^\mu \gamma_{2,\mu}$.

It is instructive to compare Eqs. \eqref{eq:integrodiff2} with the multi-time equations that Dirac originally proposed in his article "Relativistic Quantum Mechanics" \cite{dirac_32} (for two electrons):
\begin{equation}
	[ -i \gamma_k^\mu \partial_{k,\mu}- e_k \gamma_k^\mu \widehat{A}_\mu(x_k) + m_k ]\psi(x_1,x_2,\mathscr{A})=0,~~k=1,2,
	\label{eq:diracsmodel}
\end{equation}
where $\widehat{A}(x)$ is an operator-valued field acting on additional field degrees of freedom in the wave function, here schematically denoted by $\mathscr{A}$.
According to Dirac \cite[p. 459]{dirac_32}:
\begin{quote}
	The interaction of the two electrons is due to the motions of both being connected with the same field.
\end{quote}
This viewpoint was worked out in more detail by Dirac, Fock and Podolsky \cite{dfp} for a multi-time model for quantum electrodynamics with a fixed number of electrons. 
The model encounters ultraviolet (UV) divergences and is therefore only well-defined if one employs regularization procedures. These procedures, however, may have further drawbacks; introducing a momentum cutoff, for example, breaks Lorentz invariance.

Equations \eqref{eq:integrodiff2}, by contrast, constitute an action-at-a-distance version of Dirac's model \eqref{eq:diracsmodel}. First of all, in \eqref{eq:integrodiff2}, $\psi$ does not contain any degrees of freedom for the field. The field terms \eqref{eq:fieldterms} in the two equations of \eqref{eq:integrodiff2} are different and can be associated with the respective other particle. For example, $\widehat{A}_2(y)$ in the equation for particle 1 is, for every $y$, an integral operator acting solely on the degrees of freedom of particle 1. If $\psi(x_1,x_2)$ is a product wave function $\psi_1(x_1)\otimes \psi_2(x_2)$, then $\widehat{A}_2(y)$ acts only on $\psi_2$. Thus, there is a reason to say that $\widehat{A}_2(y)$ is the \textit{field term generated by particle 2}. In the first equation of \eqref{eq:integrodiff2}, $\widehat{A}_2(y)$ is evaluated at $y=x_1$, and analogously $\widehat{A}_1(y)$ at $y=x_2$ in the second equation. For \eqref{eq:integrodiff2} it therefore seems appropriate to change Dirac's quote to:
\begin{quote}
	The interaction of the two electrons is due to the motion of each being connected with the field generated by the other.
\end{quote}
This principle avoids the self-interaction which seems to be the source of the UV divergence problem, and which precludes a well-defined dynamics.

 The fact that the UV divergence is the main obstacle to a well-defined dynamics can most clearly be seen in the case of classical electrodynamics (see \cite{feynman_nobel_lecture}).
In this case, Wheeler and Feynman \cite{wf1,wf2}, building on previous work of Gau{\ss}, Fokker, Tetrode and Schwarzschild, were able to find natural and well-defined equations of motion for charged point particles interacting along light cones.
The law for the world-line $x_k(\tau_k)$ of the $k$-th particle includes a sum of advanced and retarded field tensors as follows:
\begin{equation}
	m_k \ddot{x}_k^\mu(\tau_k) = e_k \sum_{j\neq k} \tfrac{1}{2} [F_{j,+} + F_{j,-}]^{\mu \nu}\big(x_k(\tau_k)\big) \, \dot{x}_{k,\nu}(\tau_k),~~~k=1,...,N.
	\label{eq:wfeom}
\end{equation}
Here,
\begin{equation}
	F_{j,\pm}^{\mu \nu}(x) = \partial^\mu A_{j,\pm}^{\nu}(x) - \partial^\nu A_{j,\pm}^{\mu}(x),
\end{equation}
and the vector potentials are given by the retarded ($-$) and advanced (+) Li\'{e}nard-Wiechert potentials, i.e.
\begin{equation}
	A_{j,\pm}^{\mu}(x) = \frac{e_j \,\dot{x}_j^\mu(\tau_{j,\pm}(x))}{[x-x_j(\tau_{j,\pm}(x))]_\nu \, \dot{x}_j^\nu(\tau_{j,\pm}(x))},
	\label{eq:wffields}
\end{equation}
where the parameters $\tau_{j,\pm}(x)$ are implictly defined by
\begin{equation}
	(x-x_j(\tau_{j,\pm}))_\mu \, (x-x_j(\tau_{j,\pm}))^\mu = 0,~~x_j^0(\tau_{j,+})> x^0,~~x_j^0(\tau_{j,-}) < x^0,
\end{equation}
i.e., those values of the $\tau_j$-parameters of the $j$-th world line $x_j^\mu(\tau_j)$ where it intersects the past ($-$) or future (+) light cone at $x$.

The integro-differential multi-time equations \eqref{eq:integrodiff2} are analogous to the Wheeler-Feynman theory \eqref{eq:wfeom}-\eqref{eq:wffields} in many ways.
\begin{enumerate}
	\item In both cases, there are no degrees of freedom for the fields. As the fields are not dynamical, their role is reduced to being book keeping devices -- convenient mathematical representations for the direct action of one particle on another.
	\item Both in \eqref{eq:wfeom} and \eqref{eq:integrodiff2}, the only field terms that appear in the equation for particle $k$ are those generated by the other particles $j \neq k$. The fields are minimally coupled.
	\item Both \eqref{eq:wfeom}-\eqref{eq:wffields} and \eqref{eq:fieldterms} express an interaction along past and future light cones. Moreover, the two theories are time-symmetric if symmetric Green's functions are used.
	\item Neither of the two equations does have a Cauchy data initial value problem. For the WF theory, this follows from the fact that \eqref{eq:wfeom} makes use, via \eqref{eq:wffields}, of points and velocities of the other particles' world lines at the intersection with past and future light cones. For the integro-differential multi-time equations \eqref{eq:integrodiff2}, the field terms \eqref{eq:fieldterms} likewise involve integrals along past and future light cones.
\end{enumerate}
In conclusion, there are good reasons to regard \eqref{eq:integrodiff2} as a quantum analog of the WF theory. This is remarkable because the latter is a non-Hamiltonian theory and has as such been resisting (canonical) quantization.\footnote{Previous non-canonical attempts of quantization of WF electrodynamics have been made in  \cite{davies_1970,davies_1971,davies_1972,hoyle_narlikar,barut_multiparticle} and \cite[chap. 8]{dirk_phd}. See also Feynman's Nobel Lecture\cite{feynman_nobel_lecture}.}
The analogy of \eqref{eq:integrodiff2} with WF electrodynamics raises new hope that \eqref{eq:integrodiff2} avoids the self-interaction problem, i.e., the main obstacle to a well-defined dynamics, also at the quantum level. 
This, in turn, may lead to a first well-defined interacting relativistic quantum-theoretical model in 1+3 spacetime dimensions.

\section{$N$-particle generalizations} \label{sec:npart}

In this section, we turn to the question of how to obtain an $N$-particle generalization of the two-particle integral equation \eqref{eq:twopartint} and also of the integro-differential equations \eqref{eq:integrodiff}. We discuss the following two approaches:
\begin{enumerate}
	\item Taking sums of pair interaction terms (Sec. \ref{sec:npartsums}).
	\item Integral equations with $N$-particle interaction kernels (Sec. \ref{sec:npartint}).
\end{enumerate}
We again focus on the case of Dirac particles.

\subsection{Sums of pair interaction terms} \label{sec:npartsums}

One can take the sum of interaction terms at two different levels: (a) at the level of the integro-differential multi-time equations, and (b) at the level of the integral equation. These two possibilities differ. We start with (a), as we have a better physical understanding of the meaning of the interaction terms at the level of the integro-differential equations.

\paragraph{Taking sums of pair interaction terms at the level of the integro-differential equations.}

The analogy of the integro-differential multi-time equations \eqref{eq:integrodiff} with WF theory suggests a particular $N$-particle generalization. From the WF point of view, we expect that the interaction term in the multi-time equation of the $k$-th particle is given by the sum of fields terms generated by the other particles, evaluated at the location of the $k$-th particle.

It only remains to specify the field terms. In the two-particle case have we observed that the field term $\widehat{A}_k$ generated by the $k$-th particle only acts on the degrees of freedom of the wave function that are associated with that particle. Postulating that this is also true for the $N$-particle case, the field terms $\widehat{A}_k$ have the following straightforward generalization to the $N$-particle case:
\begin{equation}
	[A_{j,\mu}(y) \psi](x_1,...,x_N) = i\frac{e_j}{4\pi}\int d^4 x_j' \, S_j(x_j-x_j') \delta \big(s^2(y,x_j') \big) \gamma_{j,\mu}\psi(x_1,...,x_j',...,x_N).
\end{equation}
However, in this term, values of $\psi$ on $\M^N\backslash \overline{\spacelike}$ appear. If we insist that $\psi$ must only be defined on the natural domain $\overline{\spacelike}$, then we should exchange the previous $A_{j,\mu}(y)$ with
\begin{equation}
	[A_{j,\mu}(y) \psi](x_1,...,x_N) =i\frac{e_j}{4\pi} \int d^4 x_j' \, S_j(x_j-x_j') \delta \big(s^2(y,x_j') \big) \gamma_{j,\mu} \,\id_{\overline{\spacelike}} \, \psi(x_1,...,x_j',...,x_N),
\label{eq:npartfieldterms}
\end{equation}
where $\id_{\overline{\spacelike}}$ is the indicator function of $\overline{\spacelike}$.

This leads us to the following $N$-particle equations:
\begin{equation}
	D_k \psi(x_1,...,x_N) = e_k \gamma_k^\mu \!\! \sum_{1\leq j \leq N, j\neq k} [A_{j,\mu}(x_k) \psi](x_1,...,x_N),~~k=1,...,N.
	\label{eq:npartintegrodiff}
\end{equation}
In the two-particle case, these equations are equivalent to the integral equation \eqref{eq:twopartint}. However, for $N \geq 3$, inverting the operators $D_k = (-i \gamma_k^\mu \partial_{k,\mu} +m_k)$ in \eqref{eq:npartintegrodiff} leads to $N$ different integral equations. For example, for $N=3$, inverting $D_1$ in \eqref{eq:npartintegrodiff} for $k=1$ results in:
\begin{align}
	&\psi(x_1,x_2,x_3) ~=~ \psi^\free(x_1,x_2,x_2) \nonumber\\
&+i\frac{e_1 e_2}{4\pi} \int d^4 x_1' \, d^4 x_2'~ S_1(x_1-x_1') S_2(x_2-x_2') \gamma_1^\mu \gamma_{2,\mu} \delta(s^2(x_1',x_2')) \,\id_{\overline{\spacelike}} \, \psi(x_1',x_2',x_3)\nonumber\\
&+i\frac{e_1 e_3}{4\pi} \int d^4 x_1' \, d^4 x_3'~ S_1(x_1-x_1') S_3(x_2-x_2') \gamma_1^\mu \gamma_{3,\mu} \delta(s^2(x_1',x_3')) \,\id_{\overline{\spacelike}} \,  \psi(x_1',x_2,x_3'),
\end{align}
while inverting $D_2$ in \eqref{eq:npartintegrodiff} for $k=2$ yields:
 \begin{align}
	&\psi(x_1,x_2,x_3) = \psi^\free(x_1,x_2,x_2) \nonumber\\
&+ i\frac{e_1 e_2}{4\pi} \int d^4 x_1' \, d^4 x_2'~ S_1(x_1-x_1') S_2(x_2-x_2') \gamma_1^\mu \gamma_{2,\mu} \delta(s^2(x_1',x_2'))\,\id_{\overline{\spacelike}} \,  \psi(x_1',x_2',x_3)\nonumber\\
&+ i\frac{e_2 e_3}{4\pi} \int d^4 x_2' \, d^4 x_3'~ S_2(x_1-x_1') S_3(x_3-x_3') \gamma_2^\mu \gamma_{3,\mu} \delta(s^2(x_2',x_3')) \,\id_{\overline{\spacelike}} \, \psi(x_1,x_2',x_3').
\end{align}
These two integral equations are different. This raises doubts about the consistency of \eqref{eq:npartintegrodiff}. We therefore explore the other possibility of taking sums of pair interaction terms.

\paragraph{Taking sums of pair interaction terms at the level of the integral equation.} Considering that only values of $\psi$ on the natural domain $\overline{\spacelike}$ should occur, this leads to:
\begin{align}
	&\psi(x_1,...,x_N) ~=~ \psi^\free(x_1,...x_N)\nonumber\\
& + i \sum_{i < j}  \frac{e_i e_j}{4\pi} \! \int d^4 x_i' \, d^4 x_j'\, S_i(x_i-x_i') S_j(x_j-x_j') \gamma_i^\mu \gamma_{j,\mu} \delta(s^2(x_i',x_j')) \,\id_{\overline{\spacelike}} \, \psi(x_1,...,x_i',...,x_j',...,x_N).
	\label{eq:npartintsummed}
\end{align}
We see that interaction effects are again associated with a pair $(x_i,x_j)$ of points being light-like. In fact, similarly to the boundary integral equation for two particles \eqref{eq:twopartboundaryint}, \eqref{eq:npartintsummed} can be solved autonomously on $\partial \spacelike$ and then be used as a formula to find $\psi$ on the interior of $\spacelike$.

As \eqref{eq:npartintsummed} is just a single integral equation, the question of the existence of solutions is less problematic than for the system of equations \eqref{eq:npartintegrodiff}. 
Thus, from a mathematical perspective, \eqref{eq:npartintsummed} seems preferable to \eqref{eq:npartintegrodiff}. However, we would also like to see, at least on a heuristic level, that \eqref{eq:npartintsummed} makes sense physically. To this end, we now compare \eqref{eq:npartintsummed} with the previous guess \eqref{eq:npartintegrodiff} which has a clear physical motivation, being inspired by WF electrodynamics.

Acting on \eqref{eq:npartintsummed} with the operators $D_k$ (separately for each $k$) results in:
\begin{align}
	&D_k \psi(x_1,...,x_N) = i \sum_{j \neq k} \frac{e_j e_k}{4\pi} \int d^4 x_j'~S_j(x_j-x_j') \gamma_k^\mu \gamma_{j,\mu} \delta(s^2(x_k,x_j')) \,\id_{\overline{\spacelike}} \,\psi(...,x_j',...)\nonumber\\
&+ i \!\!\! \sum_{i < j,\, i,j \neq k}  \!\!\! \frac{e_i e_j}{4\pi} \! \int d^4 x_i' \, d^4 x_j'~ S_i(x_i-x_i') S_j(x_j-x_j') \gamma_i^\mu \gamma_{j,\mu} \delta(s^2(x_i',x_j')) D_k\,\id_{\overline{\spacelike}} \,\psi(...,x_i',...,x_j',...)
	\label{eq:npartintsummedintegrodiff}
\end{align}
for $k=1,...,N$.
The first line of \eqref{eq:npartintsummedintegrodiff} agrees with \eqref{eq:npartintegrodiff}. However, the term in the second line of \eqref{eq:npartintsummedintegrodiff} is not present in \eqref{eq:npartintegrodiff}. Because of the presence of $D_k \psi$ on the right hand side of \eqref{eq:npartintsummedintegrodiff}, one would not have guessed this system straightforwardly. Note that we can now insert \eqref{eq:npartintsummedintegrodiff} into itself on the right hand side. This shows that the term in the second line is of higher order in the coupling constants. Consider the case $e_k = e~\forall \,k$. Then that term is of order $\lambda^2 \approx 1/(137)^2 \ll 1$. Thus, the difference between \eqref{eq:npartintsummedintegrodiff} and \eqref{eq:npartintegrodiff} is only a small correction. So from our current perspective the two equations should be considered equally suitable with regard to non-relativistic physics. However, as the integral equation \eqref{eq:npartintsummed} from which \eqref{eq:npartintsummedintegrodiff} derives seems mathematically preferable to \eqref{eq:npartintegrodiff} because of the consistency issue, we are led to regard that integral equation \eqref{eq:npartintsummed} as the natural generalization of \eqref{eq:twopartint}.

We now show that \eqref{eq:npartintsummed} also has the expected non-relativistic limit.

\paragraph{Non-relativistic limit.} Based on the multi-time integro-differential equations \eqref{eq:npartintsummedintegrodiff}, one can start an analogous consideration as in Sec.\@ \ref{sec:nonrellimit}. As we are looking only for an approximate equation, we are allowed to neglect the higher order term in the second line of \eqref{eq:npartintsummedintegrodiff}. Then we are left with \eqref{eq:npartintegrodiff}. Now the terms in \eqref{eq:npartintegrodiff} are  similar to the ones in the two-particle equation. The only substantial difference is that now the indicator function $\id_{\overline{\spacelike}}$ occurs in the interaction terms \eqref{eq:npartfieldterms}. However, after neglecting the time delay in the interaction terms, all points $x_1,...,x_N$ in the argument of $\psi$ become synchronized, and on simultaneous configurations we have $\id_{\overline{\spacelike}} = 1$. 
Proceeding otherwise as in Sec. \ref{sec:nonrellimit} thus results in a Schr{\"o}dinger equation with Coulomb-type pair potentials:
\be
	V(t,\vx_1,\vx_2) = \sum_{i<j} \frac{e_i e_j}{4\pi} \left( \frac{1}{|\vx_i-\vx_j|} - \frac{\boldsymbol{\alpha}_i \cdot \boldsymbol{\alpha}_j}{|\vx_i-\vx_j|} \right).
\ee

\subsection{Integral equations with $N$-particle interaction kernels} \label{sec:npartint}

Another natural possibility for an $N$-particle generalization of \eqref{eq:twopartint} are integral equations with $N$-particle kernels (here for $N$ Dirac particles):
\begin{align}
	\psi(x_1,...,x_N) &= \psi^\free(x_1,...,x_N) + \lambda \int d^4 x_1' \, \cdots d^4 x_N' ~ S_1(x_1-x_1') \cdots S_N(x_1-x_1') \nonumber\\
&~~~\times K(x_1',...,x_N') \psi(x_1',...,x_N').
\label{eq:npartintgeneral}
\end{align}
Like \eqref{eq:npartintsummed}, \eqref{eq:npartintgeneral} is again a single integral equation and as such does not seem to face any consistency problem. 

Note that \eqref{eq:npartintgeneral} includes $N$ Green's functions. This leads to $N$-particle interaction terms in the corresponding integro-differential equations (obtained by acting on \eqref{eq:npartintgeneral} with $D_k$): 
\begin{align}
	D_k \psi(x_1,...,x_N) &= \lambda \int d^4 x_1' \cdots \widehat{d^4 x_k'} \cdots d^4 x_N' ~ S_1(x_1-x_1') \cdots \widehat{S_k}(x_k-x_k') \cdots S_N(x_1-x_1') \nonumber\\
&~~~\times K(x_1',...,x_k,...,x_N') \psi(x_1',...,x_k,...,x_N'),
\label{eq:npartintegrodiffgeneral}
\end{align}
where $\widehat{(\cdot)}$ denotes omission.
Such $N$-particle interaction terms are somewhat counter-intuitive. Note, however, that the non-relativistic $N$-particle Schr\"odinger equation, when reformulated as an integral equation as in Sec. \ref{sec:derivation}, also includes a (synchronized) product of $N$ Green's functions:
\begin{align}
	\varphi(t,\vx_1,...,\vx_N) = \varphi^\free(t,\vx_1,&...,\vx_N) + \lambda \int dt' \int d^3 \vx_1' \cdots \vx_N' \, G_1(t-t',\vx_1-\vx_1') \times \cdots \nonumber\\
	&\times G_N(t-t',\vx_1-\vx_1') V(t',\vx_1',...,\vx_N') \varphi(t',\vx_1',...,\vx_N')
	\label{eq:npartschroedint}
\end{align}
with the single-time wave function \eqref{eq:singlemulti} and with
\be
	V(t,\vx_1,...,\vx_N) = \sum_{i<j} V_{ij}(t,\vx_i,\vx_j).
\ee
From this perspective, it seems natural to start with an integral equation that includes $N$ Green's functions.\footnote{Note, however, that it is not \textit{necessary} to start with an integral equation with $N$ Green's function in order to obtain \eqref{eq:npartschroedint} in the non-relativistic limit. The previous integral equation \eqref{eq:npartintsummed} also leads to a Schr\"odinger equation in differential form which is, of course, equivalent to  \eqref{eq:npartschroedint}.}

The next question is which interaction kernels are appropriate for \eqref{eq:npartintgeneral}. Recall that for the two-particle integral equation \eqref{eq:twopartint}, the interaction kernel can be thought of as arising from the natural Poincar\'{e} invariant measure on $\partial \spacelike$ (see the boundary integral equation \eqref{eq:twopartboundaryint}). This idea has a natural extension to equations of the form \eqref{eq:npartintgeneral}, as we shall explain now. In the $N$-particle case, we have:
\be
	\partial \spacelike = \bigcup_{i < j} \partial \spacelike_{ij},
\ee
where
\be
	 \partial \spacelike_{ij} = \{ (x_1,...,x_N) \in \partial \spacelike: s^2(x_i,x_j) = 0\}.
\ee
In words: $\partial \spacelike_{ij}$ is the set of configurations $(x_1,...,x_N)$ where $x_i$ and $x_j$ are light-like related and all other pairs of points $x_k, x_l$ are either space-like or light-like related.\\
$\partial \spacelike_{ij}$ carries the following natural Poincar\'{e}-invariant measure:
\be
	d \sigma_{ij}(x_1,...,x_N) = \delta(s^2(x_i,x_j)) \prod_{(k,l) \neq (i,j)} \!\!\! \theta(-s^2(x_k,x_l)) ~ d^4 x_1 \cdots d^4 x_N,
	\label{eq:npartmeasure}
\ee
where $\theta$ denotes the Heaviside function.
Let $\kappa_{ij}$ be factors depending on $ij$, for example $\kappa_{ij} = \lambda \gamma_i^\mu \gamma_{j,\mu}$.
These ingredients make it possible to write down an $N$-particle version of \eqref{eq:twopartboundaryint}:
\begin{align}
	\psi(x_1,...,x_N) = \psi^\free(x_1,...,x_N) + \sum_{i<j} \int_{\partial \spacelike_{ij}} \!\!\! &d \sigma_{ij}(x_1',...,x_N') \, S_1(x_1-x_1') \times \cdots \nonumber\\
	& \times S_N(x_N-x_N') \, \kappa_{ij} \, \psi(x_1',...,x_N').
	\label{eq:npartboundaryint}
\end{align}
This is the $N$-particle analog of the boundary integral equation \eqref{eq:twopartboundaryint}, and it appears to us that \eqref{eq:npartboundaryint} is the most natural possibility in the class of integral equations \eqref{eq:npartintgeneral}.

	However, we have not been able to see that \eqref{eq:npartboundaryint} leads to a  Schr{\"o}dinger equation with pair potentials (possibly in the integral form \eqref{eq:npartschroedint}) in the non-relativistic limit. This is despite $\partial \spacelike$ degenerates to the set of simultaneous configurations in the non-relativistic limit, and despite the decomposition of $\partial \spacelike$ into the sets $\partial \spacelike_{ij}$ with the measure $d \sigma_{ij}$ that includes the factor $\delta(s^2(x_i,x_j))$ seems to indicate how Coulomb pair potentials might emerge. The main difficulty is that \eqref{eq:npartboundaryint} contains more time integrals than \eqref{eq:npartschroedint}, and that the range of these additional time integrations depends on the other particles' space variables.

Nevertheless, if one is ready to modify \eqref{eq:npartboundaryint}, it is possible to obtain a Schr{\"o}dinger equation with pair potentials in the non-relativistic limit. We shall demonstrate this in Appendix \ref{sec:unnaturalnpartint}. However, the required modifications seem somewhat artificial. Because of this reason, and because the interaction terms in \eqref{eq:npartintsummedintegrodiff} are suggested by the analogy with WF electrodynamics (up to a small correction), we favor the approach via \eqref{eq:npartintsummed} and \eqref{eq:npartintsummedintegrodiff} over the one via \eqref{eq:npartintgeneral}.

 \section{Conclusions} \label{sec:discussion}

In this paper, we have explored a novel kind of quantum dynamics via integral equations for a multi-time wave function. The key idea is that the many time variables of the multi-time wave function make it possible to express direct interaction with time delay at the quantum level.

Starting from the formulation of the  Schr{\"o}dinger equation as an integral equation, we have derived a new covariant integral equation \eqref{eq:twopartint} for two particles that expresses direct interaction along light cones. It is structurally related to the Bethe-Salpeter equation, but differs from the latter in several important aspects. Moreover, the Bethe-Salpeter equation is usually  regarded as an effective equation (the fundamental theory being QFT) while our equation has been motivated independently. We have also shown that our equation reduces to the usual Schr\"odinger equation with a spin-dependent Coulomb potential when the time delay of the interaction is neglected.
The hope is that our integral equation can describe an intermediate level between non-relativistic QM and QFT, and taking a similar status for relativistic processes with a fixed fermion number as the  Schr\"odinger equation has in non-relativistic QM.

The integral equation \eqref{eq:twopartint} can be reformulated as a system of two integro-differential equations \eqref{eq:integrodiff}. Contrary to the usual type of multi-time equations, these equations are automatically consistent. This is true essentially because the interaction terms in the two multi-time equations \eqref{eq:integrodiff} are derived from the single integral equation \eqref{eq:twopartint}, and therefore harmonize. This is of special interest for the theory of multi-time wave functions, where usually interaction is difficult to achieve, especially for a fixed number of particles. In the case of retarded interactions, we have worked out a new consistency condition for multi-time equations with interaction terms given by integral operators \eqref{eq:scc2}, and have shown that the integro-differential equations \eqref{eq:integrodiff} satisfy this condition.

Furthermore, our equations \eqref{eq:integrodiff} have several analogies to the Wheeler-Feynman formulation of classical electromagnetism. Our work introduces fresh ideas for the program of quantizing that theory. One can entertain the hope that in the long term this could lead to a well-defined version of QFT with direct interaction instead of fields which does not require regularization and renormalization and which works for finite times instead of only for scattering processes. At present, of course, several important questions about the integral equations need to be clarified first (see Outlook).

As one step in this direction, we have addressed the question of a suitable $N$-particle generalization of our equation. Because the consistency of a single integral equation is easier to see than the consistency of a system of $N$ multi-time equations, we have searched for a suitable $N$-particle integral equation.  We have analyzed two possible routes: (a) taking sums of two-particle interaction terms and (b) using $N$-particle interaction kernels. For both routes, we have shown that there is a natural guess how the equations could look like. For (a), the analogies with WF electrodynamics together with the requirement to have a single integral equation have led to Eq. \eqref{eq:npartintsummed}. For (b), we have exploited the fact that for the two-particle integral equation, the interaction terms can be thought of as arising from the natural Poincar\'{e} invariant measure on $\partial \spacelike$, the boundary of the set of space-like configurations. The resulting boundary integral equation \eqref{eq:twopartboundaryint} has a natural generalization for $N$ particles: Eq. \eqref{eq:npartboundaryint}. Both possibilities (a) and (b) are simpler than the BS equation for $N$ particles. The latter contains $k$-particle interaction terms for $k=2,...,N$ \cite{bijtebier_bs4}. It is therefore hard to even write down the equation explicitly.

To see which of the two equations is appropriate, we have analyzed their non-relativistic limits. For possibility (a), neglecting the time delay of the interaction yields a Sch\"odinger equation with Coulomb-type pair potentials. However, we have been unable to obtain this limiting behavior for (b). Only further modifications of the equation lead to the expected limit (see Appendix \ref{sec:unnaturalnpartint}). However, these modifications seem rather artificial. Because of this, and because possibility (a) is more closely related to WF electrodynamics, we favor (a), and regard equation \eqref{eq:npartintsummed} as the more natural $N$-particle generalization of \eqref{eq:twopartint}.

\paragraph{Outlook.} Our work opens up several interesting research questions.
\begin{enumerate}
	\item \textit{Rigorous proofs of the existence of solutions.} In Sec. \ref{sec:evollaw} we have mentioned scenarios when the integral equation \eqref{eq:twopartint} has a unique solution for every solution $\psi^\free$ of the free multi-time equations. It would be desirable to show that the conditions of these scenarios can, indeed, be met. The task is simplest in the case of retarded Green's functions where the interaction terms depend only on the past. Using simplified interaction kernels, we have recently been able to obtain some first rigorous results for this case (see \cite{mtve}). While much work remains to be done to cover more realistic cases, the results in \cite{mtve} already show that the basic scheme of multi-time integral equations works well mathematically.
	\item \textit{Classical limit.} In order to obtain more than an analogies with WF electrodynamics, it would be desirable to study a suitable classical limit of Eq. \eqref{eq:twopartint} and compare it with WF electrodynamics.
	\item \textit{Conservation laws.} Our integral equation contains what one would call  "non-local interaction terms". As a consequence, it does not admit local conservation laws. One encounters a similar situation for WF electrodynamics (see \cite{wf2}). There, Wheeler and Feynman have shown that it is possible to find non-local conservation laws replacing the local ones. It would be interesting to perform a similar analysis for our integral equation. Perhaps methods developed for non-local field theories could be used to make progress \cite{marnelius,pauli_53,kegeles_oriti_2016}.
	\item \textit{Explicit solutions.} It would be desirable to find explicit solutions of the multi-time integral equations in simple cases. Some inspiration on how to do this might come from the literature about the Bethe-Salpeter equation. There, explicit solutions have been found (see \cite{green57}) for a simplified model, the Wick-Cutkosky model \cite{wick_cutkosky}. However, instead of looking for solutions with different total energies, an eigenvalue in the coupling constant $\lambda$ is studied. Therefore, the physical significance of the solutions is unclear. One would have to improve on this aspect, as well as transfer the methods to the integral equation in this paper.
\item \textit{Particle creation and annihilation.} In this paper, we have set aside particle creation and annihilation. A description of these phenomena would, of course, be required if the approach via direct interactions along light cones should ultimately lead to an action-at-a-distance reformulation of QED.  There are (at least) two possibilities: (a) to introduce a Fock space multi-time wave function (see \cite{qftmultitime}) and to consider coupled integral equations for each $N$, (b) to use a Dirac sea type argument where the Dirac sea is described via an integral equation for $N\gg 1$. Note that the stability of the Dirac sea would perhaps be easier to motivate in a theory with direct interactions than usual, as then energy can only be transferred between particles and consequently cannot be radiated away to infinity.
\end{enumerate}

\paragraph{Acknowledgments.}
I would like to thank Detlef D{\"u}rr, Lukas Nickel, Markus N\"oth and Stefan Teufel for helpful discussions. Special thanks go to Sheldon Goldstein for detailed feedback on an earlier version of the manuscript and to Roderich Tumulka for helpful discussions and suggestions, in particular with regard to the super consistency condition (Appendix \ref{sec:ccret}).
Furthermore, I would like to thank two anonymous referees for their constructive remarks.
\\[1mm]
\begin{minipage}{15mm}
\includegraphics[width=13mm]{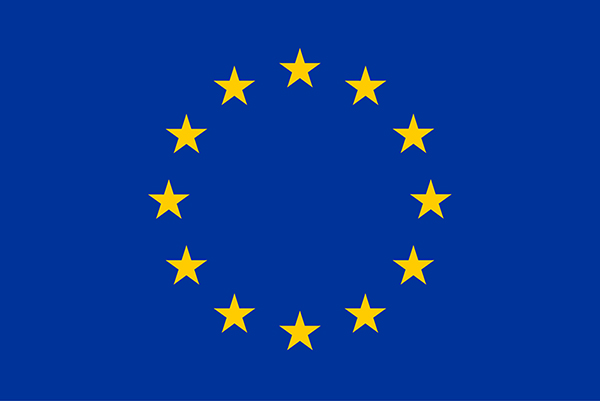}
\end{minipage}
\begin{minipage}{143mm}
This project has received funding from the European Union's Framework for Re-\\
search and Innovation Horizon 2020 (2014--2020) under the Marie Sk{\l}odowska-
\end{minipage}\\[1mm]
Curie Grant Agreement No.~705295.\\[0.2cm]
This document is the Author's Accepted Manuscript. The final published version has can be found under DOI \url{https://doi.org/10.1088/1751-8121/aae0c4} (J.\@ Phys.\@ A: Math.\@ Theor.\@ 51, 435302 (2018)). Copyright regulations by IOP Publishing apply.

\appendix
\section{Green's functions} \label{sec:greensfns}

We set $c=\hbar = 1$ and use the spacetime metric $\eta = \diag(1,\underbrace{-1,...,-1}_{d})$. Moreover, for $x \in \R^{1+d}$ we employ the shorthand notation  $x^2 = (x^0)^2-|\vx|^2$.

\paragraph{Klein-Gordon equation.} Green's functions $G_{\rm KG}$ of the (1+$d$)-dimensional Klein-Gordon equation are defined by:
\be
	(\square + m^2 )G_{\rm KG}(x) = \delta^{(1+d)}(x).
\ee
We call the time-symmetric Green's function $G^\sym_{\rm KG}$. In the various dimensions, it is given by (see \cite[chap. 7.4]{zauderer}, \cite[p. 496]{birula_qed}):
\begin{enumerate}
	\item[] $d=1$: $G^\sym_{\rm KG}(x) = \frac{1}{2} \theta(x^2) J_0(m \sqrt{x^2})$,
	\item[] $d=2$: $G^\sym_{\rm KG}(x) = \frac{1}{2\pi} \theta(x^2) \frac{\cos(m\sqrt{x^2})}{\sqrt{x^2}}$,
	\item[] $d=3$: $G^\sym_{\rm KG}(x) = \frac{1}{4\pi} \delta(x^2) - \frac{m}{8\pi \sqrt{x^2}} \theta(x^2) J_1(m\sqrt{x^2})$.
\end{enumerate}
Here, $J_0$ and $J_1$ are Bessel functions of the first kind. $\theta$ denotes the Heaviside function.

Given the time-symmetric Green's function, we obtain the retarded and advanced Green's functions as follows:
\be
	G^\ret_{\rm KG}(x) = \theta(x^0) G^\sym_{\rm KG}(x),~~~~~G^\adv_{\rm KG}(x) = \theta(-x^0) G^\sym_{\rm KG}(x).
\ee
We then have:
\be
	G^\sym_{\rm KG}(x) = \frac{1}{2} \big( G^\ret_{\rm KG}(x) + G^\adv_{\rm KG}(x)\big),
\ee
as well as $G^\sym_{\rm KG}(t,\vx) = G^\sym_{\rm KG}(-t,\vx)$.

Another important Green's function is the Feynman propagator, here for $d=3$ \cite[p. 496]{birula_qed}:
\be
	\Delta_F(x) = \frac{1}{4\pi} \delta(x^2) - \frac{m}{8\pi} \frac{\theta(x^2)}{\sqrt{x^2}} H_1^{(2)}(m\sqrt{x^2}) +  \frac{m}{8\pi} \frac{\theta(-x^2)}{\sqrt{-x^2}} K_1(m\sqrt{-x^2}),
\ee
where $H_1^{(2)}$ is a Hankel function and $K_1$ a modified Bessel function. One can see from this expression that $\Delta_F(x)$ is also time-symmetric.

Note that $\Delta(x) = 0$ for $x^2 < 0$ while this is not the case for $\Delta_F(x)$.

\paragraph{Dirac equation.} Green's functions of the Dirac equation are defined by:
\be
	\big( -i \gamma^\mu \partial_u + m\big) S_{\rm Dirac}(x) = \delta^{(1+d)}(x)\, \id.
\ee
Given a Green's function of the KG equation, one obtains a Green's function of the Dirac equation as follows:
\be
	 S_{\rm Dirac}(x) = \big( i\gamma^\mu \partial_\mu + m \big) G_{\rm KG}(x),
\label{eq:diracpropagators}
\ee
which can easily be verified by acting with $(-i \gamma^\mu \partial_u + m)$ on $G_{\rm Dirac}(x)$ and using the Clifford relations $\gamma^\mu \gamma^\nu + \gamma^\mu \gamma^\nu = 2 \eta^{\mu \nu}\, \id$.

We define the symmetric Dirac Green's function by
\be
	S^\sym(x) = \big(- i\gamma^\mu \partial_\mu + m \big) G^\sym_{\rm KG}(x) = \frac{1}{2} \big( S^\ret(x) + S^\adv(x) \big),
\ee
where $S^\ret, S^\adv$ are obtained from $G^\ret_{\rm KG}, G^\adv_{\rm KG}$ via \eqref{eq:diracpropagators}. "Symmetric" here does not mean $S^\sym(t,\vx) = S^\sym(-t,\vx)$ but rather (considering the transformation behavior of Dirac spinors under time reversal\footnote{Under time reversal, we have $\psi(t,\vx) \rightarrow \eta_T B \psi^*(-t,\vx)$, $|\eta_T|=1$.}):
\be
	B [S^\sym(-t,\vx)]^* B^{-1} = S^\sym(t,\vx)
\label{eq:symmetrytimereversal}
\ee
for the matrix $B$ defined by $B((\gamma^0)^*,(\gamma^j)^*)B^{-1} = (\gamma^0,\gamma^j)$. Here, $(\cdot)^*$ denotes complex conjugation (without transposition).

Furthermore, the Dirac Feynman propagator is given by:
\be
	S_F(x) = \big( i\gamma^\mu \partial_\mu + m \big) \Delta_F(x).
\ee
It is also time-symmetric in the sense of \eqref{eq:symmetrytimereversal}.

By integrating the relation $(-i \gamma^\mu \partial_\mu + m) S_{\rm Dirac}(x) = \delta^{(4)}(x)$ over a small time interval $-\varepsilon < x^0 < \varepsilon$, one can see that every Green's function of the Dirac equation satisfies
\be
	\lim_{\varepsilon \rightarrow 0} \big( S_{\rm Dirac}(\varepsilon,\vx) -  S_{\rm Dirac}(-\varepsilon,\vx) \big) = i \gamma^0 \delta^{(3)}(\vx).
\ee
Considering $S^\ret(t,\vx)=0$ for $t<0$, we obtain
\be
	\lim_{\varepsilon \rightarrow 0} S^\ret(\varepsilon,\vx) = i \gamma^0 \delta^{(3)}(\vx)
\label{eq:deltapropertydirac}
\ee
and analogously we find $\lim_{\varepsilon \rightarrow 0} S^\adv(-\varepsilon,\vx) = -i \gamma^0 \delta^{(3)}(\vx)$, hence $\lim_{\varepsilon \rightarrow 0} S^\sym(\varepsilon,\vx) = \frac{i}{2} \, \gamma^0 \delta^{(3)}(\vx) = -\lim_{\varepsilon \rightarrow 0} S^\sym(-\varepsilon,\vx)$.

\section{Consistency condition in the retarded case} \label{sec:ccret}

In this appendix, we derive a consistency condition for certain types of multi-time integro-differential equations where the interaction terms depend only on the past.

We consider multi-time equations
\be
	\partial_{t_k} \psi = - i (H_k^\free + I_k) \psi,~~k=1,2,
	\label{eq:multitimehamilt}
\ee
where the operators $H_k^\free$ are time-independent and commute, and the $I_k$ are linear integral operators depending only on the past. For concreteness, we consider integral operators of a similar structure as the ones in \eqref{eq:integrodiff}:
\begin{align}
	(I_1 \psi)(t_1,\vx_1,t_2,\vx_2) = \int_{-\infty}^{t_2} dt_2' \int d^3 \vx_2' ~K_1(t_1,\vx_1,t_2,\vx_2;t_2',\vx_2') \psi(t_1,\vx_1,t_2',\vx_2'),\nonumber\\
	(I_2 \psi)(t_1,\vx_1,t_2,\vx_2) = \int_{-\infty}^{t_1} dt_1' \int d^3 \vx_1' ~K_2(t_1,\vx_1,t_2,\vx_2;t_1',\vx_1') \psi(t_1',\vx_1',t_2,\vx_2).
	\label{eq:integraloperators}
\end{align}
Here, we assume that the kernels $K_1,K_2$ are differentiable, or at least that the expressions $\partial_{t_1} (I_2 \psi)$ and $\partial_{t_2} (I_1 \psi)$ do not contain any time derivatives of $\psi$. (The latter is the case for the interaction terms of \eqref{eq:integrodiff} which include distributions and hence do not have conventional derivatives).

We would like to see whether the system \eqref{eq:multitimehamilt} is consistent in the following sense: for every prescribed function $\chi$ up to $t_1,t_2 \leq t_0$ for some $t_0$, there is a function $\psi$ such that $\psi$ agrees with $\chi$ for  $t_1,t_2 \leq t_0$ and $\psi$ solves \eqref{eq:multitimehamilt} on the infinitesimal stretch $t_0 \leq t_1,t_2 \leq t_0 + dt$. If such a $\psi$ exists, we call it a \textit{conditional solution} of \eqref{eq:multitimehamilt} with history $\chi$.

Assume that such a conditional solution $\psi$ exists and is smooth. Then we can expand it into a Taylor series in the times around $t_1 = t_0 = t_1$. Let $0 \leq dt_1, dt_2 \leq dt$. Then, suppressing the spatial arguments, we have:
\begin{align}
	&\psi(t_0+dt_1,t_0+dt_2) = \psi(t_0,t_0) + (\partial_{t_1} \psi)(t_0,t_0) dt_1 +  (\partial_{t_2} \psi)(t_0,t_0) dt_2\nonumber\\
&+ (\partial_{t_1} \partial_{t_2}\psi)(t_0,t_0) dt_1 dt_2 + \frac{1}{2} (\partial_{t_1}^2 \psi)(t_0,t_0) dt_1^2 +\frac{1}{2} (\partial_{t_2}^2 \psi)(t_0,t_0) dt_2^2 + \mathcal{O}(dt^3).
\label{eq:taylor}
\end{align}
Conversely, if we can define $\psi$ unambiguously via its Taylor series and the equations of motion using only past data, we obtain a conditional solution on the infinitesimal stretch $t_0 \leq t_1,t_2 \leq t_0 + dt$. Here we shall do this up to (and including) second order in $dt$. (In this way, one can also derive the usual consistency condition \eqref{eq:cc}.) Because the expressions are lengthy, we only report on the result of the calculation.

It is indeed possible to use \eqref{eq:multitimehamilt} to express \eqref{eq:taylor} only via past data, as the interaction terms $I_k$ depend only on past data and the right hand sides of \eqref{eq:multitimehamilt} do not contain any time derivatives. 
However, the result could depend on the order in which one expresses the mixed partial derivatives $\partial_{t_1} \partial_{t_2} \psi$ via the equations of motion.
Thus, we obtain as a condition that this order must not matter. On the one hand, we have:
\begin{align}
	(\partial_{t_1} \partial_{t_2}\psi)(t_0,t_0) &= \{ \partial_{t_1} [-i(H_2^\free + I_2) \psi] \}(t_0,t_0)\nonumber\\
&=  (-i H_2^\free \partial_{t_1} \psi)(t_0,t_0) -i \partial_{t_1} (I_2 \psi)(t_0,t_0)\nonumber\\
&=  - [H_2^\free(H_1^\free + I_1) \psi](t_0,t_0) -i [\partial_{t_1} (I_2 \psi)](t_0,t_0).
\label{eq:partialt1t2}
\end{align}
We consider $-i \partial_{t_1} (I_2 \psi)(t_0,t_0)$ separately to make sure that we express all time derivatives of $\psi$ via the equations of motion. For the integral operator $I_2$ of \eqref{eq:integraloperators} we obtain:
\begin{align}
	[\partial_{t_1} (I_2 \psi)](t_0,\vx_1,t_0,\vx_2) &= \int d^3 \vx_2' K_2(t_0,\vx_1,t_0,\vx_2;t_0,\vx_2') \psi(t_0,\vx_1,t_2,\vx_2')\nonumber\\
&+ \int_{-\infty}^{t_1} dt_1' \int d^3 \vx_1' ~(\partial_{t_1} K_2)(t_0,\vx_1,t_0,\vx_2;t_1',\vx_1') \psi(t_1',\vx_1',t_0,\vx_2).
\label{eq:partiall2}
\end{align}
The crucial point is: according to our assumptions on $K_2$ there is not any term containing $\partial_{t_1} \psi$ remaining. So there is no need to express any further time derivatives of $\psi$ in \eqref{eq:partialt1t2} via the equations of motion.\footnote{In the case that the $I_k$ are multiplication operators, the situation is different. In that case, there is a further time derivative of $\psi$ to be expressed via the equations of motion and the above consideration leads to the usual consistency condition \eqref{eq:cc}.}
 We understand from now on that $\partial_{t_1} (I_2 \psi)$ is shorthand for the expression \eqref{eq:partiall2}.

On the other hand, we find:
\begin{align}
	(\partial_{t_2} \partial_{t_1}\psi)(t_0,t_0) &= \{ \partial_{t_2} [-i(H_1^\free + I_1) \psi] \}(t_0,t_0)\nonumber\\
&=  (-i H_1^\free \partial_{t_2} \psi)(t_0,t_0) -i \partial_{t_2} (I_1 \psi)(t_0,t_0)\nonumber\\
&=  - [H_1^\free(H_2^\free + I_2) \psi](t_0,t_0) -i [\partial_{t_2} (I_1 \psi)](t_0,t_0).
\label{eq:partialt2t1}
\end{align}
So the consistency condition reads:
\be
	 [H_2^\free(H_1^\free + I_1) \psi](t_0,t_0) +i [\partial_{t_1} (I_2 \psi)](t_0,t_0) = [H_1^\free(H_2^\free + I_2) \psi](t_0,t_0) +i [\partial_{t_2} (I_1 \psi)](t_0,t_0),
\ee
which can be simplified to (recalling $[H_1^\free,H_2^\free] = 0$):
\be
	[(i\partial_{t_1} - H_1^\free)I_2 \psi](t_0,t_0) = [(i\partial_{t_2} - H_2^\free)I_1 \psi](t_0,t_0).
	\label{eq:scc2}
\ee
As $\psi$ has to agree with arbitrary $\chi$'s for $t_1,t_2 \leq t_0$, \eqref{eq:scc2} must hold for arbitrary functions $\psi$. Furthermore, also $t_0$ is arbitrary. So the condition \eqref{eq:scc2} really is a condition only for the operators $H_1^\free, H_2^\free$ and $I_1,I_2$. As we demand the compatibility with arbitrary histories $\chi$, not only $\chi$ which satisfy the equations of motion in the past, condition \eqref{eq:scc2} may be stronger than needed just for the existence of (some) solutions. Therefore, we call \eqref{eq:scc2} the \textit{super consistency condition} (SCC).

\paragraph{The SCC applied to the integro-differential equations \eqref{eq:integrodiff}.}
For $H_k^\free = H_k^{\rm Dirac}$ and $I_k = \gamma_k^0 L_k$, where the $L_k$ are operators only involving the past, \eqref{eq:scc2} can be recast as
	\be
		[D_1 (L_2 \psi)](t_0,t_0) = [D_2 (L_1 \psi)](t_0,t_0).
	\label{eq:scc3}
	\ee
 Consider the case that $L_1, L_2$ are the interaction terms in the multi-time integro-differential equations that result from acting with $D_1, D_2$ on the general integral equation \eqref{eq:twopartintgeneral} with retarded Green's functions, i.e.:
\begin{align}
	(L_1 \psi)(x_1,x_2) &= \int d^4 x_2' ~ G_2^\ret(x_2-x_2') \, K(x_1,x_2') \, \psi(x_1,x_2'),\nonumber\\
	(L_2 \psi)(x_1,x_2) &= \int d^4 x_1' ~ G_1^\ret(x_1-x_1') \, K(x_1',x_2) \, \psi(x_1',x_2).
\end{align}
This includes \eqref{eq:integrodiff} as a special case.
Then, for all $\psi$, we have:
\be
	D_1 (L_2 \psi) = K(x_1,x_2) = D_2(L_1 \psi),
\ee
so the SCC is satisfied. We summarize: interaction terms derived from an integral equation of the form \eqref{eq:twopartintgeneral} automatically satisfy the SCC. So also from the perspective of demanding the existence of conditional solutions for arbitrary histories, integral equations are a good starting point.

\section{An integral equation with $N$-particle interaction kernel that has a plausible non-relativistic limit} \label{sec:unnaturalnpartint}

In the following, we show that a suitable modification of the integral equation \eqref{eq:npartboundaryint} leads to a Schr\"odinger equation with Coulomb-type pair potentials.

We focus on the case of $N$ Dirac particles and consider the integral equation \eqref{eq:npartintgeneral} with interaction kernel
\be
	K(x_1,...,x_N) = \lambda \sum_{i<j} K_{ij}(x_1,...,x_N),
\ee
where
\be
	K_{ij}(x_1,...,x_N) = \gamma_i^\mu \gamma_{j,\mu} \,  \delta(s^2(x_i,x_j)) \!\! \prod_{k \notin \{ i,j\}} \frac{1}{2\pi} \left[ \frac{\theta(-s^2(x_i,x_k))}{\sqrt{-s^2(x_i,x_k)}} + \frac{\theta(-s^2(x_j,x_k))}{\sqrt{-s^2(x_j,x_k)}} \right].
	\label{eq:specialpairkernel}
\ee
$K$ is a manifestly Poincar\'{e} invariant interaction kernel. However, contrary to \eqref{eq:npartboundaryint}, it only includes Heaviside functions which force each $x_k$ with $k \notin \{ i,j\}$ to be space-like related to either $x_i$ or $x_j$ but not to both. Moreover, no further spacelike relations between the $x_k$ with $k \notin \{ i,j\}$ (in case there several such $k$) are enforced. As a consequence, the integral equation then involves configurations outside of the natural domain $\overline{\spacelike}$. Nevertheless, the special product of Heaviside functions together with the weight functions $(-s^2(\cdots))^{-1/2}$ does allow us to obtain great simplifications in the non-relativistic limit, as we shall see now.

We compare the integral equation \eqref{eq:npartintgeneral} with interaction kernel given by \eqref{eq:specialpairkernel} with the integral formulation of the single-time Schr\"odinger equation \eqref{eq:npartschroedint}. The main assumption is similar to the one in Sec. \ref{sec:nonrellimit}, namely that we are allowed to neglect time delay in integrals over the multi-time wave function.
Moreover, we shall make use of the identity
\be
	I := \frac{1}{\pi} \int d x_k^0 ~ \frac{\theta(-s^2(x_i,x_k))}{\sqrt{-s^2(x_i,x_k)}} =  1,
	\label{eq:identity}
\ee
which we shall demonstrate now.
\be
	I~\stackrel{\tau = x_k^0-x_i^0}{=}~ \frac{1}{\pi} \int_{-|\vx_i-\vx_k|}^{|\vx_i-\vx_k|} \frac{d \tau}{\sqrt{|\vx_i-\vx_k|^2 - \tau^2}}
	~=~ \frac{1}{\pi} \left. \arcsin \left( \frac{\tau}{|\vx_i-\vx_k|} \right) \right|_{-|\vx_i-\vx_k|}^{|\vx_i-\vx_k|}	~=~ 1.
\ee
Let us turn begin with the limiting consideration. The integral equation \eqref{eq:npartintgeneral} with \eqref{eq:specialpairkernel} has the structure
\begin{align}
	\psi(x_1,...,x_N) = \psi^\free(x_1,...,x_N) + \lambda \sum_{i<j} \int d^4 x_1' &\cdots d^4 x_N' \, S_1(x_1-x_1') \cdots S_N(x_N-x_N')\nonumber\\
	&\times K_{ij}(x_1',...,x_N') \psi(x_1',...,x_N').
	\label{eq:npartintspecialkernel}
\end{align}
Consider a summand with index $ij$, i.e.
\begin{align}
	\int & d^4 x_1' \cdots d^4 x_N' \, S_1(x_1-x_1') \cdots S_N(x_N-x_N')  \,  \gamma_i^\mu  \gamma_{j,\mu} \, \delta(s^2(x_i',x_j')) \nonumber\\
	&\times  \left( \prod_{k \notin \{ i,j\}} \frac{1}{2\pi} \left[ \frac{\theta(-s^2(x_i',x_k'))}{\sqrt{-s^2(x_i',x_k')}} + \frac{\theta(-s^2(x_j',x_k'))}{\sqrt{-s^2(x_j',x_k')}} \right] \right) \psi(x_1',...,x_N').
	\label{eq:ijterm}
\end{align}
Rearranging yields:
\begin{align}
	\eqref{eq:ijterm} &= \int d^4 x_i' \, d^4 x_j' \, S_i(x_i-x_i') S_j(x_j-x_j') \,   \gamma_i^\mu \gamma_{j,\mu} \, \delta \big(s^2(x_i',x_j') \big)\, \times \nonumber\\
	& \times \left\{ \prod_{k \notin \{ i,j\}} \left( \int d^4 x_k' \, S_k(x_k-x_k') \,
	\frac{1}{2\pi} \left[ \frac{\theta(-s^2(x_i',x_k'))}{\sqrt{-s^2(x_i',x_k')}} + \frac{\theta(-s^2(x_j',x_k'))}{\sqrt{-s^2(x_j',x_k')}} \right] \right) \right\} \psi(x_1',...,x_N').
	\label{eq:eqwithproduct}
\end{align}
Now focus on one of the nested integrals in the product over $k$, and on one of the two summands in the inner bracket. Neglecting time delay, we treat $S_k \psi$ as constant in the time interval $t_k' \in \big[t_i'-|\vx_k'-\vx_i'|,~ t_i' + |\vx_k'-\vx_i'|\big]$. This yields:
\begin{align}
	& \int d^4 x_k' \, S_k(x_k-x_k') \, 
	\frac{1}{2\pi} \frac{\theta(-s^2(x_i',x_k'))}{\sqrt{-s^2(x_i',x_k')}} \, \psi(x_1',...,x_N')\nonumber\\
	\approx &\int  d^3 \vx_k' \, S_k(t_k-t_i',\vx_k-\vx_k') \, \psi(x_1',...,t_k'=t_i',\vx_k',...,x_N') \int d t_k' \, \frac{1}{2\pi} \frac{\theta(-s^2(x_i',x_k'))}{\sqrt{-s^2(x_i',x_k')}}.
	\label{eq:limitcalcint0}
\end{align}
This puts us into the position to use \eqref{eq:identity} for the $t_k'$-integral. We find:
\be
	\eqref{eq:limitcalcint0} = \frac{1}{2} \int d^3 \vx_k' \, S_k(t_k-t_i',\vx_k-\vx_k')\, \psi(x_1',...,t_k'=t_i',\vx_k',...,x_N').
\ee
Proceeding in this way for all $k \notin \{i,j \}$, we obtain $2^{N-2}$ summands resulting from the product in \eqref{eq:eqwithproduct}, of the form
\begin{align}
	&\int d^4 x_i'  \, d^4 x_j' \, S_i(x_i-x_i') S_j(x_j-x_j') \,  \gamma_i^0 \gamma_i^\mu \gamma_j^0 \gamma_{j,\mu} \, \delta \big(s^2(x_i',x_j') \big) \, \times \nonumber\\
	& \times \frac{1}{2^{N-2}}\left( \prod_{k \notin \{ i,j\}} \int d^3 \vx_k' \, S_k(t_k - \tau_k', \vx_k-\vx_k') \right) \psi(...,x_i',...,x_j',...,t_k'=\tau_k',\vx_k,...)
	\label{eq:limitcalcint1}
\end{align}
where $\tau_k' \in \{t_i',t_j' \}$. Next, we perform the integral over $t_j'$, using $ \delta(s^2(x_i',x_j'))$ decomposed according to \eqref{eq:deltadecomp}. After neglecting time delay another time, \eqref{eq:limitcalcint1} reduces to:
\begin{align}
	\eqref{eq:limitcalcint1} &\approx \int d^4 x_i' \, d^3 \vx_j' \, S_i(x_i-x_i') S_j(t_j - t_i',\vx_j-\vx_j') \,  \frac{\gamma_i^\mu \gamma_{j,\mu} }{|\vx_i'-\vx_j'|} \times\nonumber\\
	 \times &\frac{1}{2^{N-2}}\left( \prod_{k \notin \{ i,j\}} \int d^3 \vx_k' \, S_k(t_k - t_i', \vx_k-\vx_k') \right) \psi(...,x_i',...,t_j'=t_i',\vx_j,...,t_k'=t_i',\vx_k,...)
	\label{eq:limitcalcint2}
\end{align}
Now all the $2^{N-2}$ summands in \eqref{eq:eqwithproduct} are equal in the approximation. Relabel $t_i'$ as $t'$. Then \eqref{eq:npartintspecialkernel} evaluated at equal times $t_l = t ~\forall l$ becomes:
\begin{align}
	\psi(t,\vx_1,...,t,\vx_N) \approx ~&\psi^\free(t,\vx_1,...,t,\vx_N) + \lambda \sum_{i<j} \int d t' \, d^3 \vx_1' \cdots d^3 \vx_N' \, S_1(t-t',\vx_1-\vx_N') \times \cdots \nonumber\\
	&\times S_N(t-t',\vx_N-\vx_N') \, 
  \frac{\gamma_i^\mu \gamma_{j,\mu} }{|\vx_i'-\vx_j'|}\,  \psi(t',\vx_1',...,t',\vx_N').
\end{align}
We thus obtain the integral version \eqref{eq:npartschroedint} of the single-time Schr{\"o}dinger equation with spin-dependent Coulomb pair potentials.


\end{document}